\documentclass{aa}  
\usepackage{hyperref}
\hypersetup{colorlinks=true,
            allcolors=blue}
\usepackage{graphicx}
\usepackage{booktabs}
\usepackage{wasysym} 
\usepackage{txfonts}
\usepackage{float}
\bibpunct{(}{)}{;}{a}{}{,}

\def\fpres{$f_{\rm pres}$}
\def\fpresT{$\Tilde{f}_{\rm pres}$}

\begin{document}

   \title{Effects of planetary day-night temperature gradients on He~1083~nm transit spectra}

   \author{F. Nail \inst{1},
            A. Oklopčić \inst{1},
            M. MacLeod \inst{2}
          }

   \institute{Anton Pannekoek Institute for Astronomy, University of Amsterdam, 1090 GE Amsterdam, Netherlands
   \and 
   Center for Astrophysics, Harvard \& Smithsonian 60 Garden Street, MS-16, Cambridge, MA 02138, USA
            }

   \date{Received 2023 August 11; accepted 2023 December 8.}
 
\abstract{A notable fraction of helium observations probing the evaporating atmospheres of short-period gas giants at 1083~nm exhibit a blueshift during transit, which might be indicative of a day-to-night side flow. 
In this study, we explore the gas dynamic effects of day-to-night temperature contrasts on the escaping atmosphere of a tidally locked planet. Using a combination of 3D hydrodynamic simulations and radiative transfer post-processing, we modeled the transmission spectra of the metastable helium triplet.
Our key findings are as follows: (1) Increasing the day-night anisotropy leads to a narrowing of the helium line and an increase in the blueshift of the line centroid of a few km~s$^{-1}$. (2) The velocity shift of the line depends on the line-forming altitude, with higher planetary mass-loss rates causing the line to form at higher altitudes, resulting in a more pronounced velocity shift. (3) A critical point of day-night anisotropy comes about when the blueshift saturates, due to turbulent flows generated by outflow material falling back onto the planet's night side. (4) A strong stellar wind and the presence of turbulent flows may induce time variations in the velocity shift.
Assuming that the day-night temperature gradient is the main cause of the observed blueshifts in the He-1083~nm triplet, the correlation between the velocity shift and day-night anisotropy provides an opportunity to constrain the temperature gradient of the line-forming region. 
}


\maketitle

%-------------------------------------------------------------------
\section{Introduction}

%%% BIG PICTURE %%%
The escape of atmospheres is thought to play a key role in shaping the demographics of exoplanets. Significant loss of gas from planetary atmospheres has been proposed as a potential explanation for two intriguing features observed in the exoplanetary radius-orbital distance diagram: the so-called ``radius valley'' and the ``hot Neptune desert.'' The radius valley refers to the bimodal distribution of small planet sizes, suggesting a bifurcation between planets that are stripped of and retain their gaseous atmospheres \citep{lopez_role_2013, owen_kepler_2013, owen_evaporation_2017, kurokawa_mass-loss_2014, fulton_california-_2017, vaneylen_asteroseismic_2018, allan_evolution_2019, hallatt_sculpting_2022}. The hot Neptune desert denotes a lack of Neptune-sized planets in proximity to their host stars, where stellar irradiation levels are high \citep{szabo_short-period_2011, lundkvist_hot_2016, mazeh_dearth_2016}.
By studying a diverse range of exoplanetary systems, we can gain deeper insights into the process of atmospheric escape and its role in shaping the evolution of exoplanets.

The absorption of helium at 1083~nm \citep[e.g.,][]{ nortmann_ground-based_2018, salz_detection_2018, kirk_kecknirspec_2022}, the hydrogen Lyman-$\alpha$ line \citep[e.g.,][]{vidal-madjar_extended_2003, ehrenreich_giant_2015}, and various UV lines of metal species \citep[e.g.,][]{sing_hubble_2019, cubillos_near-ultraviolet_2020, yan_detection_2021, sreejith_cute_2023} observed during planetary transits serve as evidence of atmospheric escape. These spectral features probe the low-density, outflowing upper atmosphere of exoplanets. Among them, the helium feature at 1083~nm has emerged as an excellent tracer due to its favorable properties, which includes being less susceptible to the effects of the interstellar medium (ISM) and more accessible with ground-based instruments \citep[e.g.,][]{oklopcic_new_2018}.

For planets that are relatively close to their host stars, tides might act to tidally lock their rotation to their orbital motion. In this case, a single side of the planet would face the star at all times. Over time, energy input from the stellar continuum or the extreme ultraviolet (XUV) radiation of the host star could shape a contrast between the day and night side temperatures of the planet. The value of the temperature likely depends on whether the atmospheric loss is powered by photoevaporation due to XUV irradiation \citep{lammer_atmospheric_2003, murray-clay_atmospheric_2009}, in which case the day-side temperature might reach $\sim 10^4$~K for typical parameters or by the gravitational energy of the planetary core  \citep{ginzburg_core-powered_2018, gupta_sculpting_2019}; in this case, a lower temperature that is closer to the planet's equilibrium temperature, given the stellar bolometric luminosity, would be predicted  \citep[e.g.,][]{bean_nature_2021}. Regardless of the mechanism, a temperature (and thus pressure) gradient from day to night side of a planetary atmosphere would drive hydrodynamic rearrangement of those layers in potentially observable flows.

Indeed, some observations of the helium 1083~nm line have shown evidence for a net blueshift of the line at mid-transit, which would correspond to a bulk flow away from the star toward the observer
\citep{nortmann_ground-based_2018, salz_detection_2018, alonso-floriano_he_2019, allart_spectrally_2018, kirk_confirmation_2020, palle_he_2020, zhang_detection_2023}. The goal of this work is to study the gas dynamical effects of a day-to-night temperature contrast on flows in the vicinity of a planet with an escaping atmosphere. We can then model how various atmospheric properties lead to observable differences in the transit spectra by generating synthetic helium 1083~nm spectra.

Thus far, several approaches have been employed to model planetary outflows, as well as to predict and interpret their signatures in the helium line \citep[e.g.,][]{allart_spectrally_2018, shaikhislamov_global_2020, wang_metastable_2021, wang_metastable_2021-1, khodachenko_impact_2021, macleod_stellar_2022, schreyer_using_2023}. Among these, the hydrodynamics of anisotropic escaping atmospheres was previously studied by \citet{stone_anisotropic_2009,villarreal_dangelo_sensitivity_2014, carroll-nellenback_hot_2017,wang_metastable_2021, wang_metastable_2021-1}. The present work builds on these prior studies by adopting the parameterized 3D hydrodynamic approach from \cite{macleod_stellar_2022}. 
The paper is structured as follows. In Sect\,\ref{sec:Methods&Models}, we describe our numerical models to simulate the anisotropic planetary wind. In Sect\,\ref{sec:Results}, we present the results of the hydrodynamic simulations and their radiative transfer analysis. In Sect\,\ref{sec:Discussion}, we discuss the implications of our results for the interpretation of helium observations of short-period gas giants regarding their day-night anisotropy and possible observing strategies. We conclude with a summary in Sect\,\ref{sec:Summary}.

%--------------------------------------------------------------------
\section{Methods and models} \label{sec:Methods&Models}

%%% INTRO METHODS %%%
We investigated the interaction between the thermal winds of a planet and its host star, while varying the degree of day-night anisotropy of the planet. To achieve this, we used the 3D Eulerian (magneto)hydrodynamic code \texttt{Athena++}\footnote{\url{https://github.com/PrincetonUniversity/athena}}, version 2021 \citep{stone_athena_2020}, which is a derivation of the \texttt{Athena} code \citep{stone_athena_2008}. Our objective was to obtain observational signatures of different wind configurations in the helium triplet at 1083~nm by calculating the radiative transfer of the stellar light passing through the planetary outflow in a transiting configuration. 

We parameterized the conditions at the base of the planetary wind, as opposed to self-consistently simulating atmospheric heating and wind launching. We examined the observational signatures arising from different geometries of the planetary outflow. This approach has allowed us to remain agnostic with respect to the actual wind-driving mechanisms, while considering a broad range of wind properties and focusing on their observable signatures. 

This study is a continuation of the work by \cite{macleod_stellar_2022}; however, instead of assuming an initially-isotropic planetary wind, we have considered the impact of a day-night anisotropy on the planet (or, more precisely, at the base of the wind). Our focus is on identifying observational features in the helium line that are indicative of an anisotropic day-night side structure. Since we have only altered the planetary wind, while utilizing the same simulation setup and radiative transfer code presented in their work, in the following section, we  outline the main simulation features and discuss the changes made for this study. For a detailed description of the simulation, we refer to Section 2 of \cite{macleod_stellar_2022}.

%%% SIMULATION %%%
\subsection{Simulation}

We ran hydrodynamic simulations to solve the equations for mass, momentum, and energy conservation of an inviscid gas. The simulations use a spherical polar mesh with the host star at the center of the frame of reference that co-rotates with the orbital motion of the planet. The computational domain spans the star's surface to a distance of 0.3 au in all directions. The base mesh of the simulation is made up of $9 \times 6 \times 12$ meshblocks, and each block consists of $16^3$ zones. These zones are spaced logarithmically in the $r$ direction and evenly in the $\theta$ and $\varphi$ directions to maintain nearly cubic zone shapes throughout the volume. To avoid zones with extreme aspect ratios, the number of effective zones in the $\varphi$ direction is reduced near the poles. Instead of restructuring the mesh itself, this is accomplished by averaging conserved quantities across these zones. An additional five levels of static mesh refinement $N_{\mathrm{SMR}} = 5$ are used around the planet, within a box covering a range of $\pm 12$ planetary radii.

The planet in the simulation is modeled after the system WASP-107b \citep{anderson_discoveries_2017, spake_helium_2018, piaulet_wasp-107bs_2021}, with its mass and radius specified in Table \ref{tab:Model_parameter}. The planet is located at the negative x-axis at a distance of $a$ (the semi-major axis) from the star, and both the planet and star are assumed to be rotating with an orbital frequency of the planet of $\Omega = \Omega_{orb} = G(M_p+M_*)a^{-3}$. The angular momenta of the planet and star are both in the positive z-direction. 

We employed the ideal gas equation of state with the gas adiabatic index $\gamma =1.0001$. The behavior along adiabats is nearly isothermal, but gas in the domain includes the planetary and stellar winds with widely different temperatures. We adopted an isothermal equation of state in order to focus on outflow dynamics rather than its thermodynamics. In particular, the details of radiative heating and cooling are therefore ignored by adopting a single temperature as a crude representation of the balance of these processes. A more detailed approach could self-consistently treat this physics, but at the expense of the  simplified, parameterized approach that lets us change model properties and observe the effects on the system. 

We took into account the gravitational influence of the two bodies, while the back-reaction of the outflow on the planetary orbit and the effect of radiation pressure at 1083~nm on the gas dynamics were neglected. Each simulation was run for $3\times 10^{6}$~s, equivalent to slightly over six orbits of the planet around the star.

The density and pressure of the stellar surface are constant and are related by the hydrodynamic escape parameter $\lambda_*$, which denotes the ratio between the gravitational potential $U(R_*)$ and thermal energy $k_B T(R_*)$. Thus, the pressure on the stellar surface is determined by the density of the stellar surface, $\rho_*$, and $\lambda_*$. All our models use a value of $\lambda_* = 15$. Similarly, the pressure on the planetary surface is defined as
\begin{equation}
    P_p = \rho_p \frac{GM_p}{\gamma \lambda_p R_p}~,
    \label{eq:press_p}
\end{equation}
where $\lambda_p = \frac{GM_p}{c_s^2 R_p}$ denotes the hydrodynamic escape parameter for the planet. The unitless parameter reflects the thermal potential of the planet and thus, effectively determining the local gas temperature
\begin{equation} \label{eq:local_gas_T}
\centering
    T = \frac{1}{\lambda_p} \frac{\mu m_H}{k_B} \frac{GM_p}{R_p} .
\end{equation}

%%% ANISOTROPY %%%
\subsection{Day-night anisotropy}
In order to generate an anisotropic planetary outflow, we introduced a pressure scaling factor \fpres\ that scales the pressure across the planetary surface. The pressure was scaled to reach its maximum value at the substellar point and then decreases along the surface of the planet toward the antistellar point, where it reaches the predefined minimum value \fpresT. To achieve this pressure distribution, we can calculate \fpres\ as a function of the distance to the substellar point along the planet's surface
\begin{equation}
\centering
    f_{\rm pres} = \frac{1-\Tilde{f}_{\rm pres}}{2} \left[ \sin(\theta_p) \cdot \cos(\varphi_s - \varphi_p) + 1\right] + \Tilde{f}_{\rm pres}~,
\end{equation}
where $\varphi_p\in [0,2\pi]$ and $\theta_p \in [0,\pi]$ are the spherical coordinates centered around the planet, and the substellar point is defined as $S(\varphi_p=0, \theta_p=\pi/2)$. 

\begin{figure}
    \centering
    \includegraphics[width=.49\textwidth]{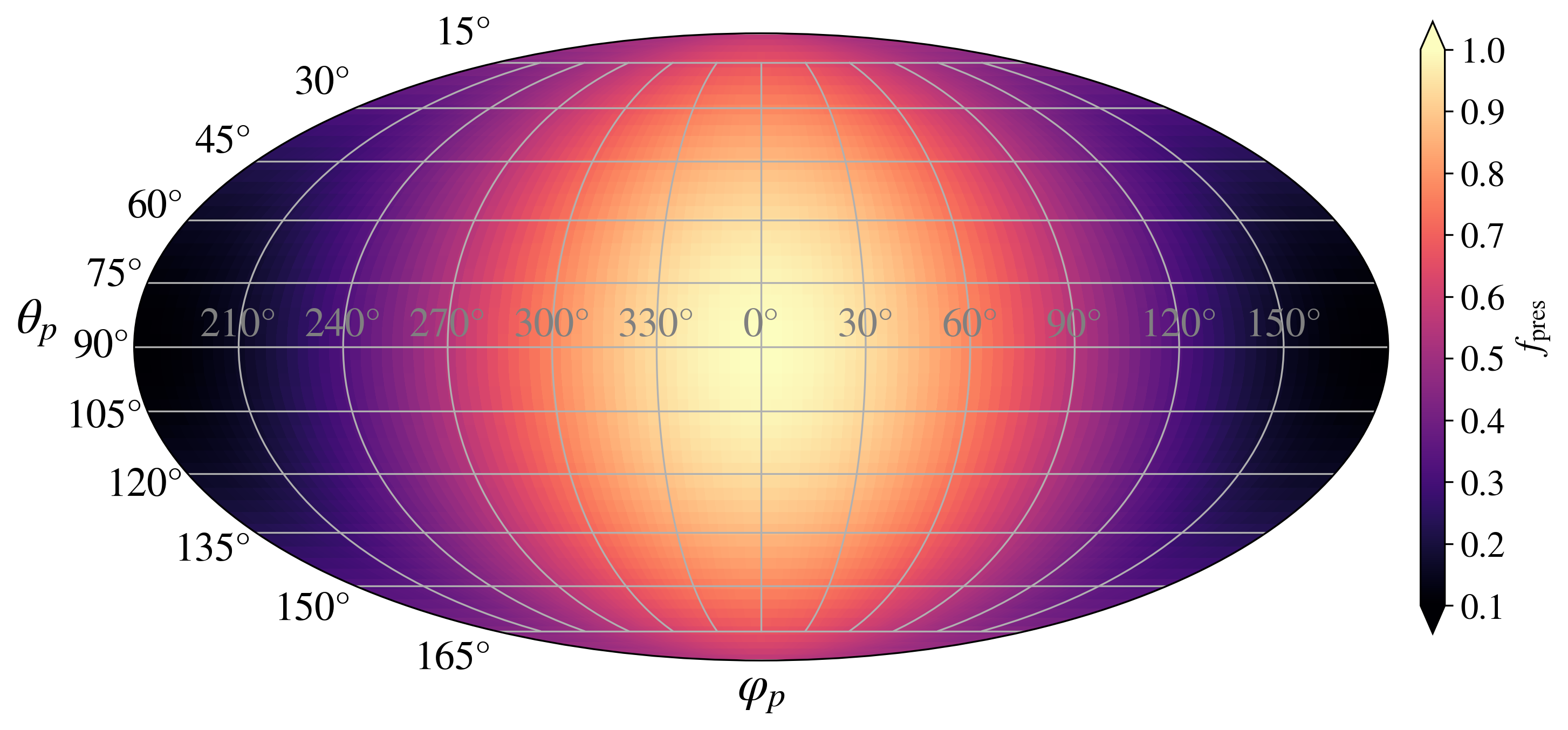}
        \includegraphics[width=0.35\textwidth]{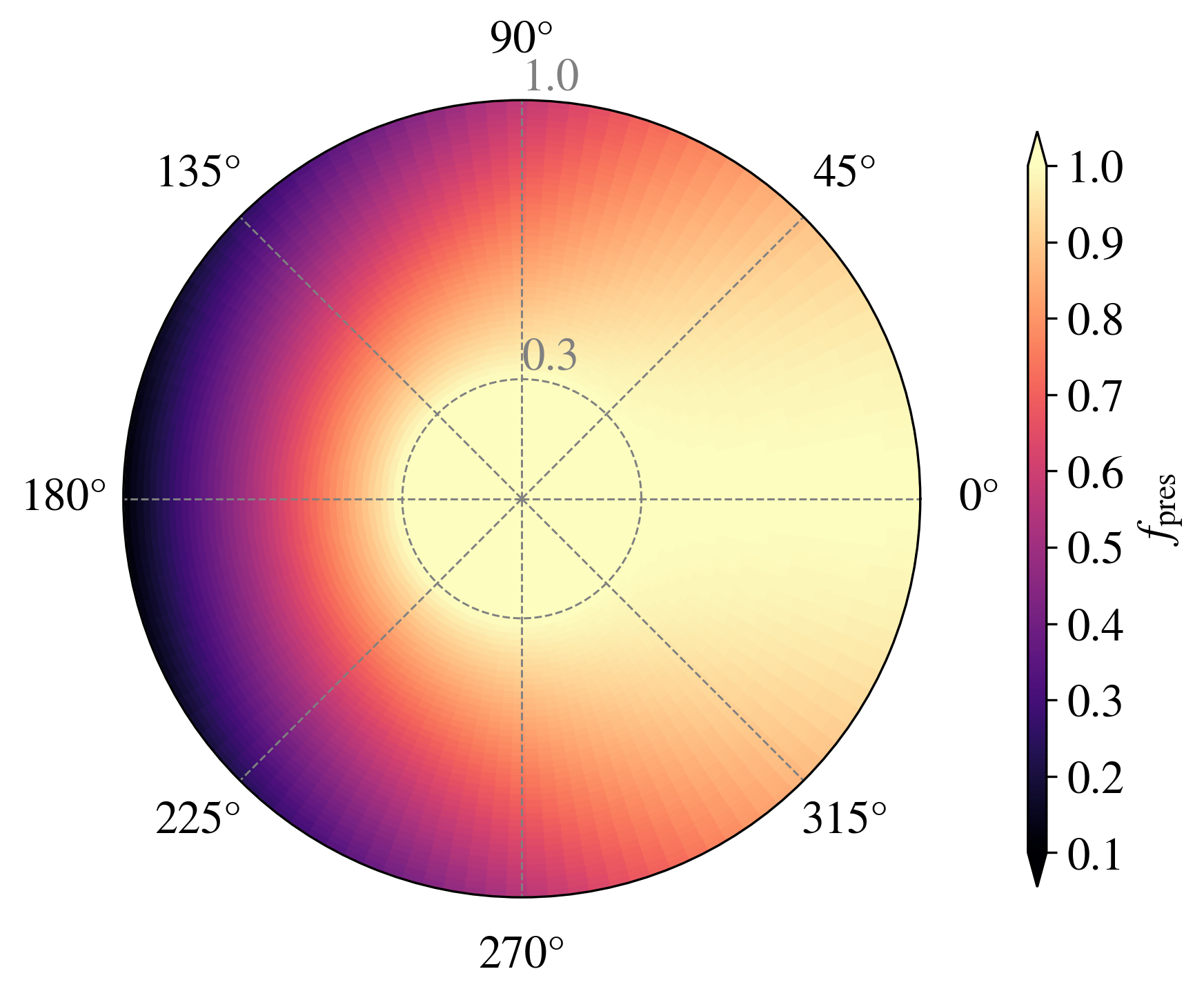}
    \caption{Maps of the pressure scaling factor of the planetary boundary condition. The upper panel shows the pressure scaling factor, \fpres\, along the planet surface, in the case where the value at the antistellar point $(\varphi_p=180^{\circ}, \theta_p=90^{\circ})$ is 10\% of the value at the substellar point $(\varphi_p=0, \theta_p=90^{\circ})$. The lower panel shows a slice through the planet, in which the observer is located in the $-x$ and the star in the $+x$-direction. For models with a lower degree of anisotropy, a similar distribution is used, only with a different minimum value on the night side \fpresT.}
    \label{fig:P_map}
\end{figure}

We varied the degree of anisotropy by adjusting \fpresT, with the maximum degree of anisotropy  achieved when the pressure at the antistellar point is only 10\% of that at the substellar point. Figure \ref{fig:P_map} displays the pressure scaling factor \fpres\ over the planet's surface and interior for this 10\% model. In addition, we modified the pressure inside the planet to initiate the wind, while avoiding discontinuities. For the interior, where $r \leq 0.3~R_p$, \fpres\ remains constant at a value of 1. For $0.3~R_p < r \leq 1~R_p$, the pressure decreases linearly toward the planetary surface to reach the pre-defined value of \fpres.

%%% WIND VARIATIONS %%%
\subsection{Wind variations}
To gain insight into the impact of each parameter and identify potential degeneracies, we analyzed a series of models. To start, we investigated a fiducial model, denoted as model $A$, which serves to illuminate the fundamental mechanisms of the anisotropic evaporative atmosphere. In this model, we used the system parameters of WASP-107b, an intermediate hydrodynamic escape parameter of $\lambda_p=2$, and a relatively weak stellar wind, with the stellar mass-loss rate only twice that of the planet. We varied only the degree of anisotropy, as indicated by the index of each model between the limits of \fpresT $= 100\%$ (isotropic), and \fpresT $=10\%$ (anisotropic). 

We then considered the effects of other model parameters on our results. We examined how a modified stellar wind affects the anisotropic outflow of the planet. This analysis is inspired by that of \cite{macleod_stellar_2022}, where the impact of intermediate (their model $B$) and strong (their model $C$) stellar winds on planetary atmospheric escape was studied. We labeled our models as $B$ and $C,$ with modified stellar mass-loss rates accordingly. Furthermore, we consider variations in the orbital distance of the planet, denoting the corresponding models with $D$. We also investigated the effects of different hydrodynamic escape parameters, modifying the value of $\lambda_p$ and labeling the corresponding models as $L$. The relevant input parameters are summarized in Table \ref{tab:Model_parameter}. It is important to highlight that the values for $\lambda_p$ represent the maximum values on the day side of the planet. However, when a day-night anisotropy is present, these maximum values decrease, along with the local gas temperature, as we move toward the night side along the planetary surface.

\begin{table}
    \caption{Model overview of the input parameters.}
    
    \begin{tabular}{lcccc}
    \toprule
         model group & $\lambda_p$ & $a$ [au] & $\rho_*$ [g~cm$^{-3}$] & $
         \rho_p$ [g~cm$^{-3}$]\\
    \midrule
        $A$ & 2 & 0.05 & $4.16 \times 10^{-15}$ & $1.53 \times 10^{-16} $\\ 
        $B$ & 2 & 0.05 & $4.16 \times 10^{-14} $ & $1.53 \times 10^{-16} $\\
        $C$ & 2 & 0.05 & $4.16 \times 10^{-13} $ & $5.21 \times 10^{-16}$\\
        $D$ & 2 & 0.03 & $4.16 \times 10^{-15}$ & $1.53 \times 10^{-16} $\\
        $L1.5$ & 1.5 & 0.05 & $4.16 \times 10^{-15}$ & $1.43 \times 10^{-16} $\\
        $L3$ & 3 & 0.05 &  $4.16 \times 10^{-15}$& $2.27 \times 10^{-16} $\\ 
    \bottomrule
        \end{tabular}
Notes. We present four groups of simulation models, categorized by the strength of the stellar wind and the planetary parameters. The system parameters are identical for all simulations, following those of WASP-107b, $M_* = 0.68~M_{\astrosun} = 1.36\times 10^{33}$~g,  $R_* = 0.67~R_{\astrosun} = 4.67\times 10^{10}$~cm,   $M_p = 0.096~M_{\jupiter}=1.82 \times 10^{29}$~g \citep{piaulet_wasp-107bs_2021}; $R_p = 0.94~R_{\jupiter} = 6.71 \times 10^{9}$~cm \citep{anderson_discoveries_2017}.\\
    \label{tab:Model_parameter}
\end{table}

%%% RADIATIVE TRANSFER %%%
\subsection{Radiative transfer and synthetic spectra}
For the radiative transfer analysis, we followed the methodology of \citet{macleod_stellar_2022}. We made the assumption that the gas composition throughout the computational domain is solar, with mass fractions of hydrogen, helium, and metals being X = 0.738, Y = 0.248, Z = 0.014, respectively. Moreover, we considered that the detailed equilibrium is reached in each cell of the simulation grid at a steady state. According to this method, the unattenuated photoionization rate, $\Phi,$ is derived from the stellar flux and, in combination with the optical depth, $\tau$, provides the photoionization rate of $\Phi e^{-\tau}$. The optical depth to hydrogen-ionizing photons is calculated iteratively for each cell in the simulation grid. The assumed stellar spectral energy distribution used in calculating the photoionization rates is that of a late F or early G star, following \citet{zhang_giant_2023}.

To generate synthetic helium spectra, we calculated the integrated optical depth as a function of wavelength along the rays that propagate from the stellar surface to the distant observer. We imposed a Voigt line profile, with a Gaussian component dependent on the local gas temperature, as given by Eq\,\ref{eq:local_gas_T}. We do not consider rays that cross the planet interior. Additionally, the contribution of the rays was weighted according to the stellar quadratic limb darkening law.

%-----------------------------------------------------------------
\section{Results} \label{sec:Results}

%%% INTRODUCTION FIDUCIAL MODEL %%%
\subsection{Fiducial model group}\label{sec:fiducial}
\begin{figure}
    \centering
    \includegraphics[width=0.48\textwidth]{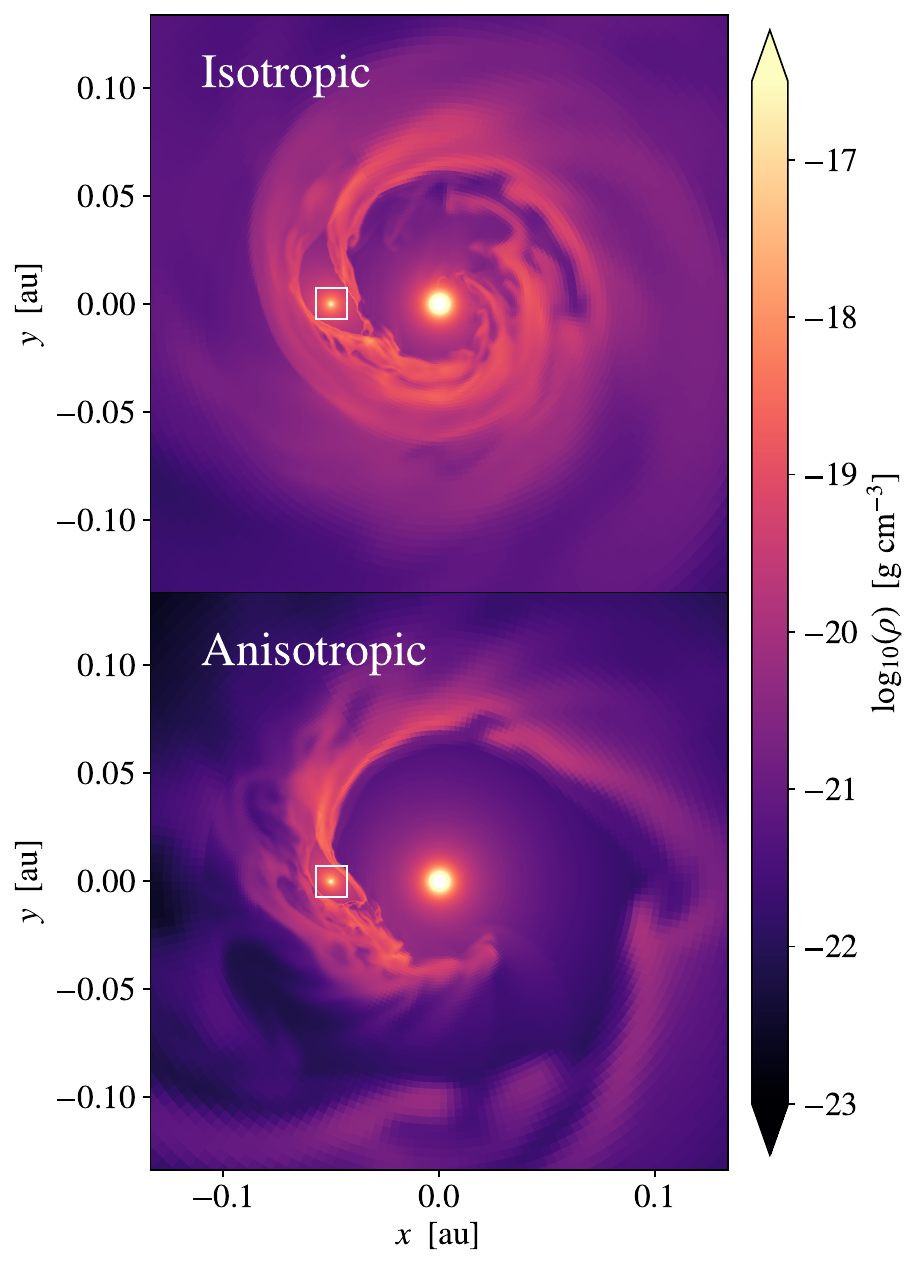}
    \caption{Gas density in the orbital midplane, showing the hydrodynamic winds generated by a model exoplanet interacting with the stellar wind emanating from the central star. We assume that the observer is located in the direction -x. The white square marks the region around the planet shown in the following figures. The top figure illustrates a planetary wind with isotropic boundary conditions ($A_{100\%}$), while the bottom figure incorporates day-night anisotropy ($A_{10\%}$). In the anisotropic scenario, the decreased pressure on the night side leads to a reduction in the mass-loss rate of the planet. As a result, the outflow becomes less extended compared to the isotropic case.}
    \label{fig:FullFrame}
\end{figure}

We started our analysis by investigating the fiducial model group, $A$. Figure \ref{fig:FullFrame} shows a slice of gas density through the orbital midplane of the simulations $A_{100\%}$ and $A_{10\%}$, which correspond to \fpresT=1 and \fpresT=0.1, respectively. If the planet is transiting in front of the star, the observer is positioned in the negative $x$-direction. The snapshots shown in Fig.\,\ref{fig:FullFrame} capture the simulations in approximately steady state, after roughly six orbits.

At large scales, the flow in these two simulations is broadly similar. Because the planets are exposed to a relatively weak stellar wind (see Table \ref{tab:app_models} and Appendix \ref{sec:app_mass_loss}), the planetary outflow forms a toroidal structure that surrounds the star \citep[e.g., as described by][]{macleod_stellar_2022}. In the presence of a day-night temperature gradient, the pressure on the night side decreases, leading to a reduction in the planetary mass-loss rate. As a result, the outflow becomes less extended compared the isotropic scenario. 

One difference with varying anisotropy factor, \fpresT, is the overall planetary mass-loss rate. With lower \fpresT, the overall energy of the planetary wind is lower, and the degree of mass loss is also reduced for a given density at the wind-launching height in the atmosphere. We plot and tabulate the model mass-loss rates given our parameter settings in Appendix \ref{sec:app_mass_loss}, in Table \ref{tab:app_models} and Fig.\,\ref{fig:app_mass_loss}.This difference is reflected in a slight decrease in the size of the cavity surrounding the planet with increasing degree of anisotropy. 

Our subsequent investigation was focused on the immediate vicinity of the planet, indicated by a white square. This immediate planetary vicinity is primarily responsible for the line-forming region in helium 1083~nm transit spectroscopy and bears the primary signatures of any asymmetry at the planetary outflow-launching scale.

%%% ANISOTROPY %%%
\subsubsection{Anisotropy}
\begin{figure*}
    \centering
     \includegraphics[width=.87\textwidth]{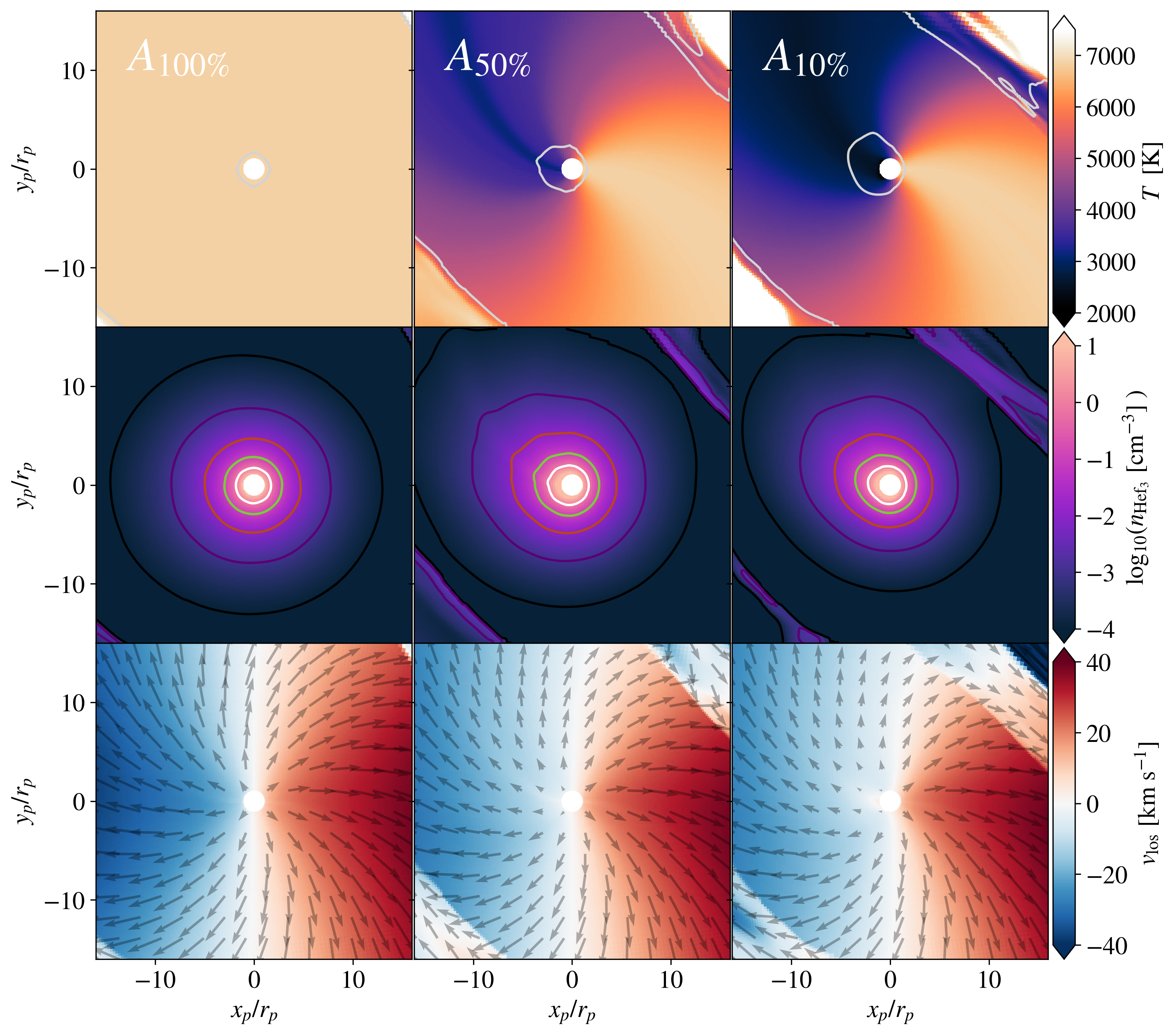}
     \caption{Temperature (top), number density of metastable helium (middle), and line-of-sight velocity (bottom) in the vicinity of the planet for increasing day-night side anisotropy from left to right. The contours in the top row are indicating the sonic surface, while the middle row displays contours of the number density of metastable helium. These contours range from light to dark and correspond to values of $n_{\rm Hef_3} = 1, 10^{-1}, 10^{-2}, 10^{-3}$, and $10^{-4}$, respectively. Velocity quivers are added in the bottom panels. The observer is located in the $-x$ direction and the star in the $+x$-direction. The plot displays the equatorial slice from a top-down perspective. By introducing anisotropy, the gas density increases toward the second quadrant (i.e., upper left) and the line-of-sight velocity becomes asymmetric around $x_p=0$.}
     \label{fig:Anisotropy}
\end{figure*}

% ISOTROPIC CASE %
Figure \ref{fig:Anisotropy} illustrates the impact of day-night side anisotropy within the unshocked region in the immediate vicinity of the planet. Starting from the left side of the figure, the model $A_{100\%}$ has an isotropic boundary condition, where the local gas temperature (top panel) remains constant throughout the atmosphere. In this isotropic state, the outflow reaches the approximately spherical sonic surface at a radius $\sim1.5~R_p$. The number density of metastable helium (middle panel) uniformly decreases radially, adhering to a Parker (spherical, isothermal) wind profile.  As a result of the nearly-radial flow, gas at $x_p >0$ generally flows toward the star (redshifted for a mid-transit observer), while gas at $x_p< 0$ generally flows away from the star (blueshifted). Due to the Coriolis force, the outflow undergoes a slight clockwise rotation, as indicated by the velocity quivers.

% ANISOTROPIC CASES %
When introducing anisotropy, we lower the pressure on the night side compared to the day side, thus reducing the gas' radial acceleration. The pressure gradient arising from this anisotropic pressure distribution acts as the driving force, causing the gas to converge toward the lower pressure region on the night side of the planet. On the night side, where the gas has accumulated, lowered pressure therefore leads to enhanced density. Additionally, as a result of the planet's orbital motion, the gas experiences a rotational component and tends to accumulate between the anti-stellar point and the evening terminator of the planet (i.e., upper left quadrant in Fig. \ref{fig:Anisotropy}). Applying the same principle, the sonic surface expands in this direction, becoming nonspherical. 

The decreased thermal energy of the night side in the anisotropic scenario causes the line-of-sight velocity to trend toward blueshift along a ray at  $x_p=0$. Material flowing toward the star (redshifted) is increasingly confined to a conical region in the $+x_p$-direction. This deviation from radial outflow reflects a net flux of gas from the day side to the night side of the planet, even as the flow remains largely outflowing.

% COMPARISON OF SONIC SURFACE TO LITERATURE %
In 2D simulations of anisotropically-launched planetary outflows, \citet{stone_anisotropic_2009} also found a distorted sonic surface. They additionally identify a shock  at $\theta_p = 3\pi / 4$, formed by the compression of the wind as it moves from the day side to the night side of the planet. In our three-dimensional (3D) models, the additional spatial degree of freedom appears to suppress this discontinuity to some extent (though see the discussion of \ref{sec:Fallback}). We observed a continuous, but distorted sonic surface in our asymmetric-heating models. Similarly, in the 3D hydrodynamic simulations by  \citet{carroll-nellenback_hot_2017, debrecht_photoevaporative_2019} outflows at the planet scale are distorted by net flux from day to night side, but do not show the broad night-side shock cone observed in the 2D result reported by \citet{stone_anisotropic_2009}.

%%% FALLBACK %%%
\subsubsection{Fallback} \label{sec:Fallback}
 \begin{figure*}
    \centering
    \includegraphics[width=.9\textwidth]{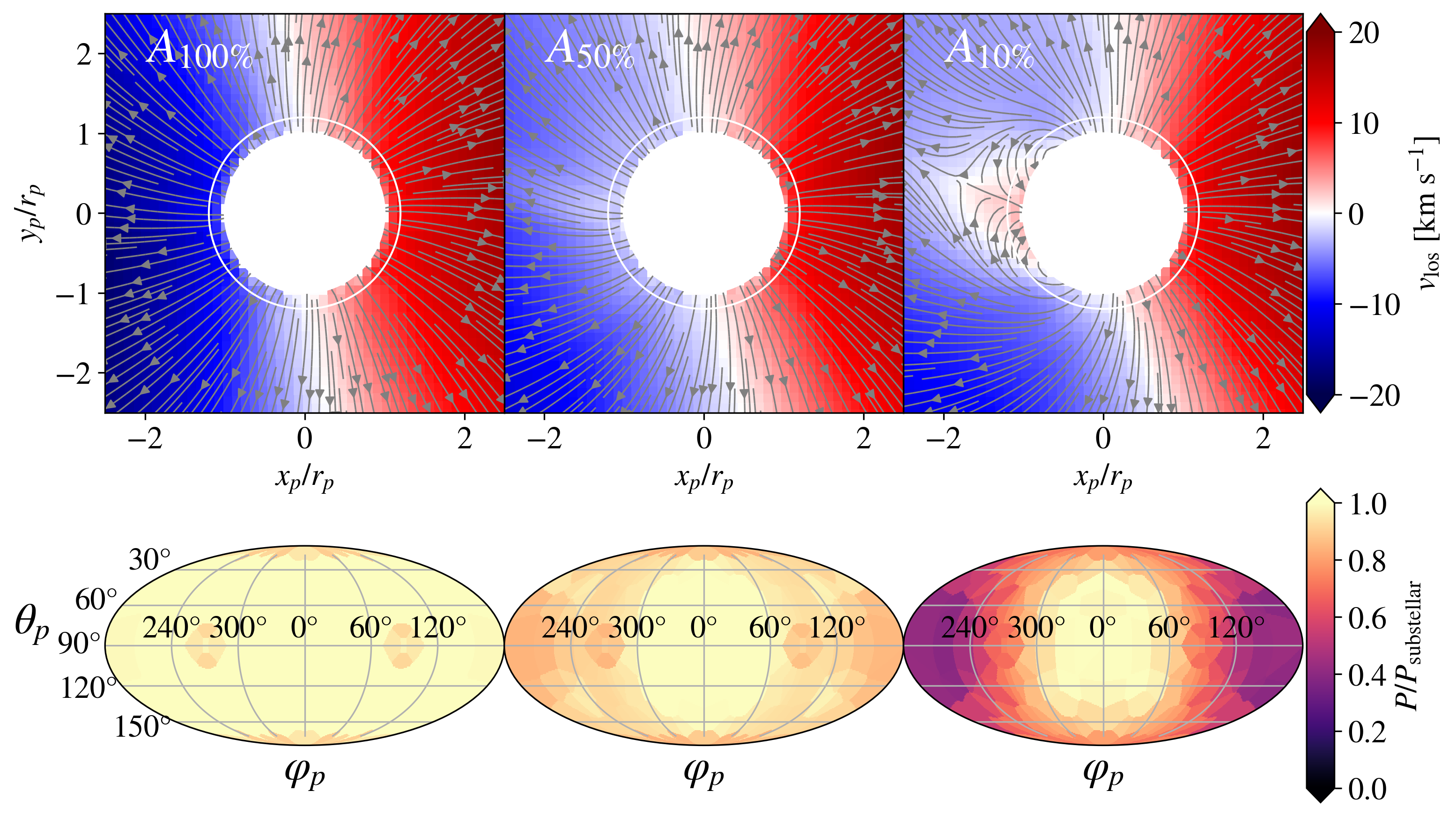}
    \caption{Line-of-sight velocity map in the close vicinity of the planet (top). The observer is located in the $-x$ and the star in the $+x$-direction. Overplotted streamlines illustrate the flow patterns and show the fallback of material on the planet's night side. The white outline indicates the radius ($r = 1.2~R_p$) of the sphere whose projected surface is shown in the maps below. Due to the incident material, the pressure on the night side in model $A_{10\%}$ reaches only 40\% instead of the expected 10\% of the pressure on the day side in the evolved simulation.
    }
    \label{fig:Fallback}
\end{figure*}

In cases where the pressure on the night side becomes very low compared to the day side, the outflowing gas on the night side stagnates and may even fall back onto the planet, as illustrated in the top right panel of Fig.\,\ref{fig:Fallback}. This fallback represents a saturation of the degree of day--night asymmetry supported by our models. For example, in the model $A_{10\%}$ in Fig.\,\ref{fig:Fallback}, higher pressure material from nearer to the day side of the planet converges on the night-side low pressure region. 
The occurrence of falling back material on the night side effectively reduces the degree of anisotropy between the day and night side. 

This saturation of a minimum night-side pressure is visualized most clearly in the pressure maps of Fig.\,\ref{fig:Fallback}. To create these surfaces, we evaluate the pressure of zones near a radius of 1.2 planetary radii, as shown by the white line in the upper panels of Fig.\,\ref{fig:Fallback}. In the $A_{100\%}$ model, the pressure is uniform, regardless of direction. In the $A_{50\%}$ and $A_{10\%}$ models, the pressure is highest in the substellar (day-side) direction, which matches conditions at the boundary condition. But the antistellar (night-side) pressure is not as low as that of the boundary. In the initial condition of model $A_{50\%}$, the pressure on the night side is set to be 50\% of the pressure on the day side. However, in the evolved steady state of the simulation, the night-side pressure increases to $\sim$80\% of the day side value. While in model $A_{10\%}$, the visible flow convergence yields an antistellar pressure of approximately $40\%$ of the substellar pressure. In each case, the net flux of material away from the day side of the planet is acting to reduce the degree of the asymmetry that has been imposed by our planetary boundary condition. 

We observe a relationship between the degree of day-night anisotropy and the opening angle of the region where the fallback occurs. We measure this angle as angular separation between points on the planet's equator where material becomes trapped and cannot escape. As the degree of day-night anisotropy increases, this fallback angular spread expands correspondingly. The occurrence of incident material is  limited to a region of about $r \leq 2~R_p$. When zoomed out of the snapshot, see for example Fig.\,\ref{fig:app_overview}, the planetary outflow appears nearly radial, due to the partial equilibration of the temperature at smaller scales. 

This finding suggests that extreme temperature gradients in exoplanetary upper atmospheres are subject to some degree of rearrangement as the material outflows from the planet. Higher degrees of anisotropy (for example at a lower atmospheric level) appear to partially equilibrate as material flows away from the planet and there is a net flux from the planet's day side toward the night side. 

We caution, however, that the exact flow morphologies we observe are not generic, but instead are likely reflections of the boundary condition we assign in our simulations (e.g. as shown in Fig.\,\ref{fig:P_map}). For example, in the more self-consistent photoionization heating models of \citet{tripathi_simulated_2015}, outflow cools day-side material, while fallback and flow convergence on the night side heat the flow in that region. These effects are not captured by our prescriptive boundary condition and simplified equation of state. For a comparison to our Fig.\,\ref{fig:Fallback}, see Figures 3, 4, and 6 of \citet{tripathi_simulated_2015}. By comparison, \citet{wang_metastable_2021,wang_metastable_2021-1} model the exoplanets WASP-69b and WASP-107b, respectively, and observe markedly cooler night-side material along with signatures of flow convergence and fallback in their models (see Figure 1 of each paper). 

The comparison of these results suggests that the exact flow morphologies, resulting temperatures, and degree of fallback to the night side are all dependent on the detailed thermodynamics and perhaps the particular planet in question. However, a general conclusion is that when the night side temperature (or pressure) becomes very low compared to the day side, there will be flow from the day side and fallback to the night side.

%%% SPECTRA %%%
\subsubsection{Synthetic Spectra}
\begin{figure*}
    \centering
    \includegraphics[width=.95\textwidth]{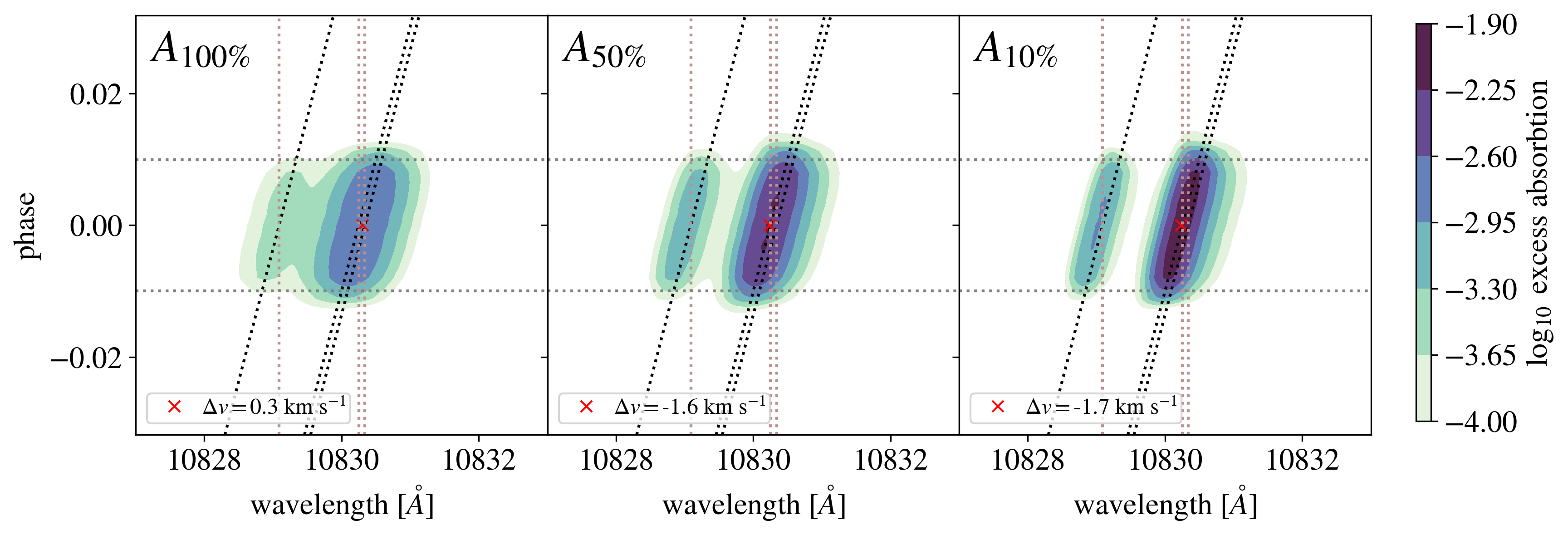}
    \caption{Excess absorption of metastable helium in the stellar frame. The degree of day-night anisotropy is increasing from left to right. The horizontal lines indicate the extent of the optical transit, while the tilted lines show the Doppler-shifted wavelengths of the helium triplet in the planetary rest frame. The vertical brown lines correspond to the stellar rest frame. The red cross indicates the wavelength of the mid-transit maximum absorption; the deviation from the vacuum wavelength is converted to a shift velocity. With an increasing day-night anisotropy, the line profiles become blueshifted and narrower. There is only minor excess absorption before and after the ingress and egress phases. }
    \label{fig:Excess}
\end{figure*}

Figure \ref{fig:Excess} presents a time series of the metastable helium excess absorption in the stellar rest frame. The synthetic spectra result from the radiative transfer analysis of simulation snapshots. To illustrate the transit time evolution, we post-process the snapshot for various observing angles. The optical transit, which is observed in the broadband spectrum, occurs within a phase range of $-0.01$ to $0.01$, equivalent to approximately 2.37 hours in total duration. %\textcolor{red}{The duration is slightly shorter due to the shorter orbital distance: 0.05 au instead of 0.055 au.}

Nonetheless, Fig.\,\ref{fig:Excess} encompasses phases preceding and following the optical transit, serving the purpose of investigating the impact of extended planetary wind material along the entire orbital trajectory. Although the outflow material occupies significant portions of the orbit, as shown in Fig.\,\ref{fig:FullFrame}, we observe only minor excess absorption pre-ingress and post-egress. However, this scenario changes for a higher density scaling (see Appendix \ref{sec:app_rescaling}), where the extended outflow leads to a notable pre- and post-transit excess absorption (i.e., Fig.\,\ref{fig:Excess_A'}).

Within the helium triplet, we can differentiate between its blue and red components, with the latter consisting of two closely spaced but unresolved lines. In observational studies, it is common to solely measure the stronger (red) component, which is why we focus on analyzing this component. In the isotropic case, the helium line is relatively broad, causing the blue and red components' wings to partially overlap. As the degree of day-night anisotropy increases, the spectral lines narrow, while their centroids shift toward shorter wavelengths. This can also be seen in the mid-transit spectra in the bottom panel of Fig.\,\ref{fig:Blueshift}.

%%% BLUESHIFT %%%
\subsubsection{Blueshift}
\begin{figure*}
    \centering
     \includegraphics[width=.9\textwidth]{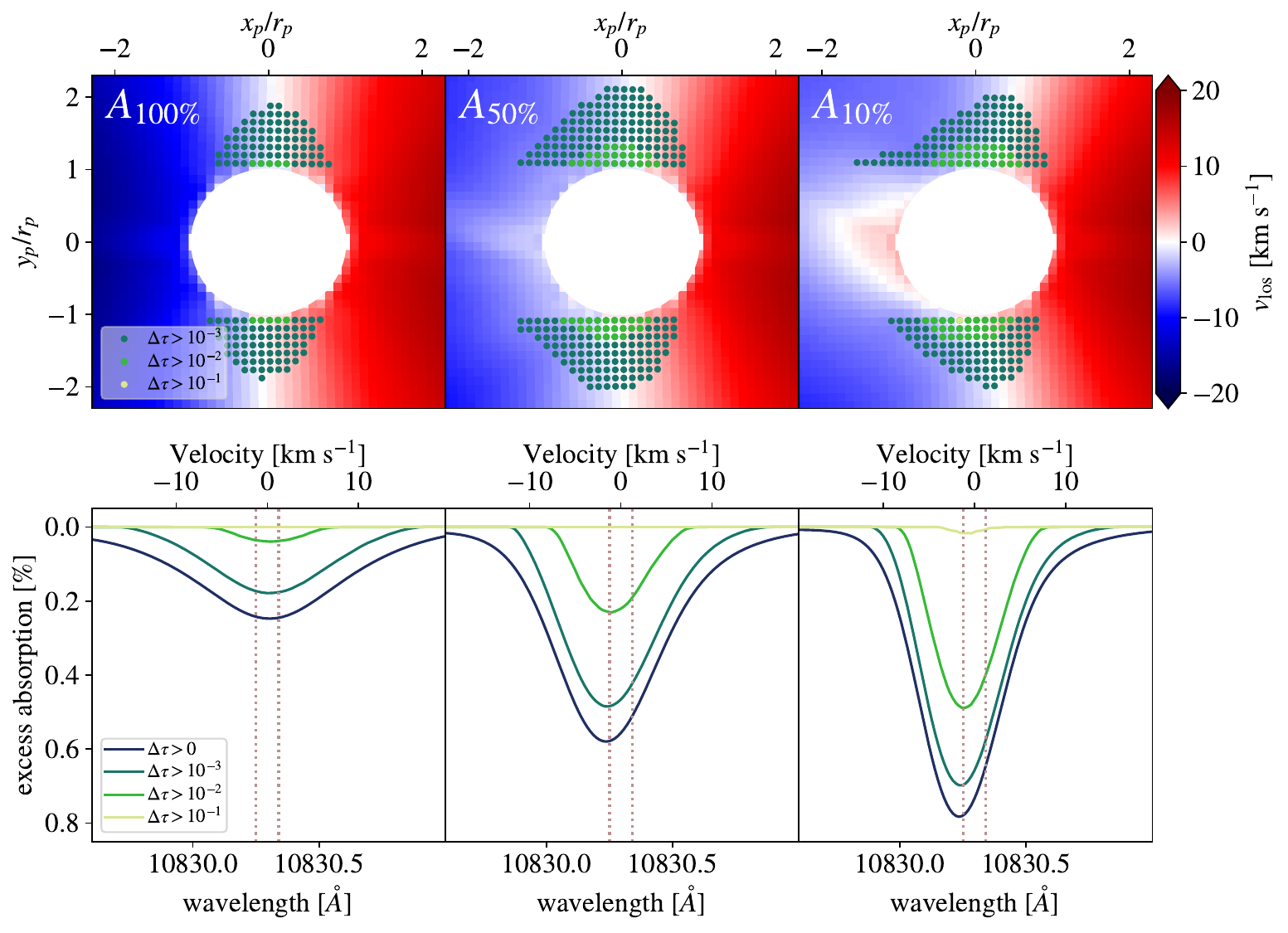}
     \caption{Zoom-in on the planet, a 2D slice of the orbital midplane, with cells indicated by optical depths greater than $10^{-3}$, $10^{-2}$, and $10^{-1}$ (from dark to light, top). The observer is located in the $-x$ and the star in the $+x$-direction. Bottom: The red component of the metastable helium triplet is calculated only from the cells that reach a certain threshold in optical depth. These spectra were computed using the full 3D snapshot. The velocity axis is centered around the peak of the red component in the stellar rest frame. This red component comprises two unresolved spectral lines, indicated here by vertical brown lines. Most of the blueshift is contributed by cells with $10^{-3} < \Delta\tau < 10^{-2}$, which corresponds to a line forming region of $r \approx 1 - 2~ R_p$.}
     \label{fig:Blueshift}
\end{figure*}

As the night-side pressure is reduced, the helium line progressively becomes blueshifted, reaching a value of $\Delta v = -1.7$~km~s$^{-1}$ for model $A_{10\%}$. 
Qualitatively, this blueshift comes from the net flow of material from the planetary day side to the night side. However, in order to investigate where in the outflow the blueshift is produced, we select individual cells close to the planet with high density in which most of the helium line is formed. To calculate the mid-transit spectra, we progressively incorporate cells with increasing optical depth $\Delta\tau$, as illustrated in the upper panels of Fig.\,\ref{fig:Blueshift}. The resulting spectra are presented in the lower panels.

In the isotropic boundary condition, velocities are symmetrically arranged around the $x_p=0$ axis. Even so, the helium line is slightly redshifted by $\Delta v = + 0.3$ km~s$^{-1}$ in the isotropic case $A_{100\%}$. We find that this is due to the finite spatial resolution of the grid and the imperfect alignment of the cells with the spherical outflow. Imperfect cancelation of positive and negative line-of-sight velocities means that a slight redshift of the resulting helium line emerges. In Appendix \ref{sec:app_numerical}, we explore how line velocity shift changes with spatial resolution during the post-processing step. Fig.\,\ref{fig:app_resolution} shows that the velocity shift for the isotropic boundary condition converges toward zero with increasing resolution level. Consequently, the redshift is  impacted by the numerical effects and the absolute numerical accuracy of the velocity shift is in the range of $\lesssim0.5$ km~s$^{-1}$. 

As the anisotropy increases, the helium line forms more in the $-x$-direction. This shift is influenced by the asymmetric density distribution caused by the net day-to-night side wind, as revealed in our prior analysis shown in Fig.\,\ref{fig:Anisotropy}. Because the mean radial velocity of the flow away from the planet is lower, the density is higher and there is an  increase in the number of cells with higher optical depth. This is particularly evident in the greater presence of light green cells in the top of Fig.\,\ref{fig:Blueshift} for models $A_{50\%}$ and $A_{10\%}$. This increase leads (as compared to the isotropic case) to a greater absorption depth of the helium line.

We note that in Figure \ref{fig:Blueshift}, the line of zero line-of-sight velocity shifts toward the star in the asymmetric cases. More of the line-forming zones have line-of-sight velocity directed toward the observer, though the typical magnitude of the velocity is lower (as can be seen in the lighter colors on the colorscale). Consequently, the resulting spectrum exhibits a net blueshift. The blueshift is amplified when considering cells with smaller optical depths. The majority of the blueshift arises from cells satisfying the condition $10^{-3} < \Delta\tau < 10^{-2}$, corresponding to a radial distance of $1-2~R_p$. As anisotropy increases, the line profile undergoes a shift toward shorter wavelengths and becomes narrower. Importantly, this narrowing is not linked to a reduction in the planetary mass-loss rate. Even when the mass-loss rate matches that of the isotropic case, as shown in the left panel of Fig.\,\ref{fig:const_mdotp}, the line remains narrow. Instead, this phenomenon can be attributed to the narrowing of the velocity distribution of the line-forming area. The maximum of the velocity distribution is reached in the blueshifted regime. At the same time, a higher absorption depth is achieved, which leads to a narrow line profile.

The discussion above highlights the importance of the radiative transfer and optical depth in shaping a helium 1083~nm line profile. 
If we increase the density of the planetary outflow, as discussed in Appendix \ref{sec:app_rescaling}, we can model the effect of a higher planetary mass-loss rate. For example, if the density is a factor of 13 higher, the line forming region can extend up to $7~R_p$, and the regions below $1~R_p$ close to the planetary surface can become opaque. This result aligns with the findings of \cite{linssen_expanding_2023}, who utilize the photoionization code Cloudy. They predict that the helium line forms between $2.5-6~R_p$ for a Neptune-like planet with a mass-loss rate similar to $10^{10}$ g~s$^{-1}$, orbiting a K-type star. 

To determine whether a portion of the lower atmosphere is optically thick, it is useful to measure the line ratios of the helium triplet. As discussed by \cite{wang_metastable_2021}, the line ratio deviates from the quantum degeneracy ratio if most of the absorption occurs in a region of higher density. According to their self-consistent photochemistry models, the line forming region of metastable helium in the atmosphere of the Saturn-mass gas giant WASP-69b is situated above the base of the photoevaporative outflow at $1.2~R_p$. Within the range of $1.9-3.8~R_p$, they observed a blueshift of $2-3$ km~s$^{-1}$, of similar magnitude to that found in our models. They attribute this blueshift to the ongoing day-night advection of the tidally locked planet, resulting from the temperature gradient between the day and night sides of the region below. 

According to our models, there is a direct correlation between the blueshift and the altitude at which the line forms. Specifically, as the line forms at higher altitudes in the escaping atmosphere, the magnitude of the blueshift increases due to higher wind speeds. This dependence on altitude is, in turn, dependent  on the density of the outflow, which is directly related to the mass-loss rate of the planet. Thus, the blueshift of the helium line is influenced by two factors: day-night anisotropy and the altitude of line formation.

Nevertheless, we consider the line velocity shift to be a valuable diagnostic tool for understanding planetary wind kinematics, as it is less vulnerable to the influence of other factors. While anisotropy does influence various aspects of the line, including depth, width, and skewness, we argue that the blueshift is a metric that is more easily determined in observed spectra. Consequently, our emphasis in this paper is directed toward exploring this feature.

%%%%%%%% PARAMETRIC STUDY %%%%%%%%%%%

%%% INTRO SHIFT VELOCITIES %%%
\subsection{Parametric study} \label{sec:Morphology}
\begin{figure}[]
    \centering
    \includegraphics[width=0.49\textwidth]{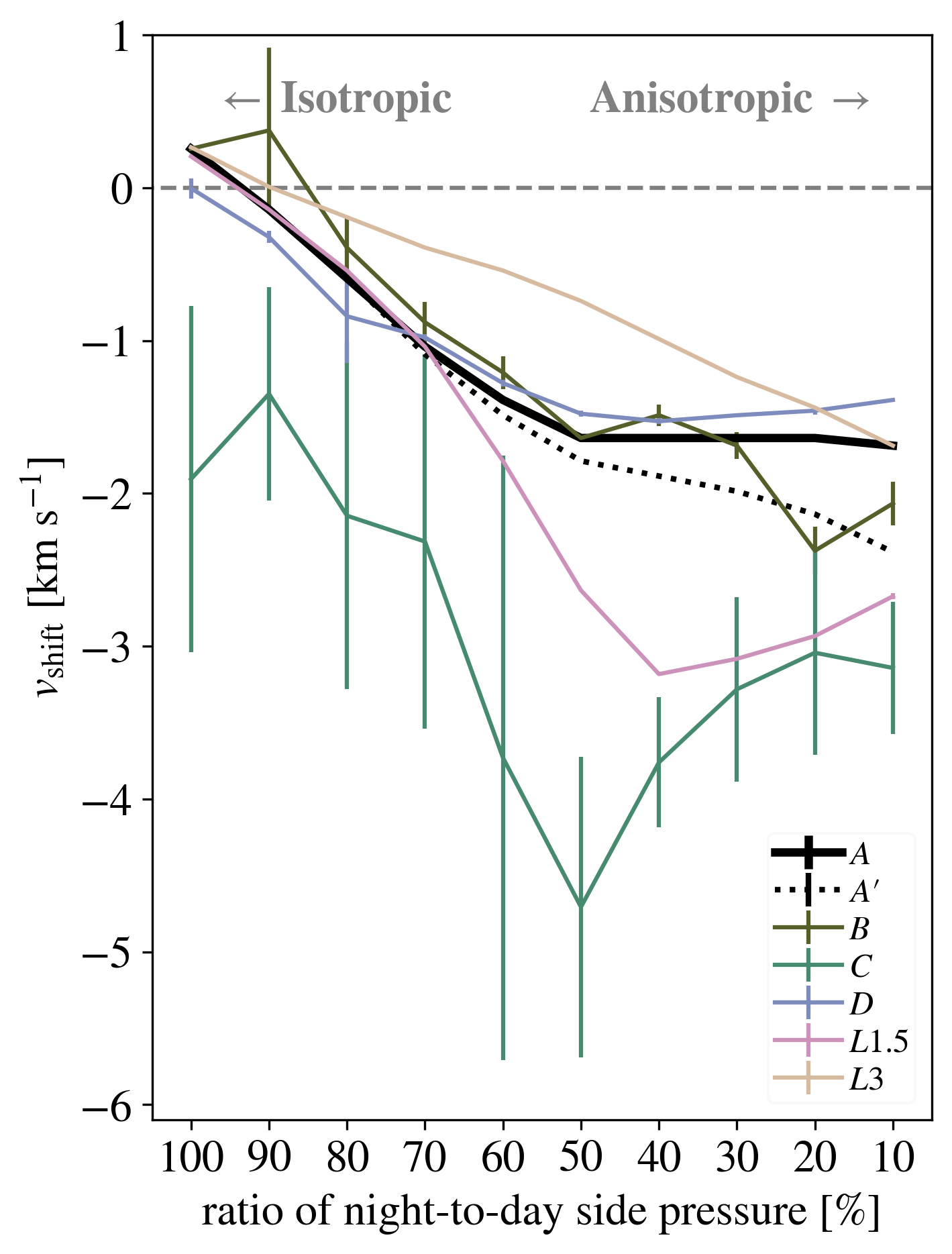}
    \caption{Mid-transit velocity shift of the dominant (red) component in the helium 1083~nm triplet as a function of the night-to-day side pressure ratio. The data points represent the average velocity shift over five snapshots in the steady-state, with error bars indicating the corresponding standard deviation. All models show the same trend: the line becomes more blueshifted with increasing day-night anisotropy. However, for ratios $< 50\%$, this behavior shifts, due to the occurrence of incident material on the night side. $A$ represents the fiducial model group, including $A'$, with the distinction that the latter employs a constant $\dot{m}_p$ instead of a constant $\rho_p$ as outlined in Sect\,\ref{sec:const_m_d_p} (see full model overview in Table \ref{tab:Model_parameter}).}
    \label{fig:Vel_shift}
\end{figure}

\noindent In this section, we aim to examine the influence of the stellar wind, orbital distance, and the planetary hydrodynamic escape parameter on the velocity shift of the helium line. Additional line features, including absorption depth and width, are detailed in Table \ref{tab:app_models}. Across all model groups, a consistent trend emerges: as day-night anisotropy intensifies, the line narrows, and absorption depth increases.

Figure \ref{fig:Vel_shift} shows the mid-transit velocity shift of the helium main component as a function of the ratio of night-to-day side pressure. A notable trend of increasing the net blueshift with increasing the level of anisotropy is observed for most models. However, the blueshift of most of the model groups reaches a plateau after an intermediate degree of anisotropy ($\lesssim 50\%$). A change in slope around this value in Fig.\,\ref{fig:Vel_shift} is related to the onset of the material fallback on the night side. As previously discussed in Fig.\,\ref{fig:Fallback}, despite setting the initial ratio of night-to-day side pressure to 10\%, it effectively reaches $\sim 40\%$ during the simulation's evolution. This indicates that the anisotropy, and consequently the net blueshift, has reached a saturated state. An overview of the outflow structure of the most anisotropic cases for each model group we consider is shown in Figure \ref{fig:app_overview}. 

%%% STELLAR WIND %%%
\subsubsection{Stellar wind}

In this section, we examine the models with increased stellar wind mass-loss rate relative to a fixed planetary mass-loss rate. While our fiducial model represents the least important stellar wind, cases $B$ and $C$ show progressively strengthening stellar wind shaping of the planetary outflow. 

In the scenario of an intermediate stellar wind, case $B$, the cavity surrounding the planet experiences a higher compression as compared to case $A$. Consequently, time variations in the blueshift of the helium line (shown in Fig.\,\ref{fig:Vel_shift} as error bars) become more prominent due to the presence of turbulent, shocked gas within in the line forming regions. However, when comparing it to case $A$ (which represents a weak stellar wind scenario), there is a small difference in magnitude of the helium line velocity shift. 

Moving on to the scenario of a strong stellar wind, denoted as case $C$, the interaction between the intense stellar wind and the planetary wind results in the suppression of the cavity surrounding the planet. This occurs due to the compression of the wind-wind interaction region, bringing it into close proximity with the planetary surface. Consequently, the outward flow originating from the planet is redirected around the planet before it reaches its sonic radius, giving rise to the formation of a tail-like structure. 

In contrast to other cases, where boundary layers between the winds were observed, in case $C$, the majority of the planetary outflow is found to be blueshifted and swept up within the compressed tail, giving rise to Rayleigh-Taylor and Kelvin-Helmholtz instabilities. As a result, the magnitude of blueshift and the variability over time are much larger compared to the other models. With a significant increase in the day-night contrast, the pressure on the night side is no longer sufficient to counteract the force of the stellar wind. Consequently, the cavity surrounding the planet on the night side dissipates, leaving only a bow shock on the day side and turbulent eddies on the night side (see Fig.\,\ref{fig:app_movC}).

%%% ORBITAL DISTANCE %%%
\subsubsection{Orbital distance}
Model $D$ is characterized by a smaller orbital distance relative to model $A$. The orbital shear velocity at this distance is higher (we emphasize that this is sometimes called the Coriolis acceleration in the context of the corotating reference frame). This shear leads to a stretched shape and contraction of the cavity around the planet, as shown in Table \ref{tab:app_models}. Consequently, higher pre- and post-transit absorptions occur with material compressed into denser streams. Notably, the helium line in model $D$ displays a Doppler-broadened helium line profile compared to case $A$. Nonetheless, the correlation between the helium blueshift and the degree of day-night anisotropy follows a similar trend in both models.

%%% HYDRODYNMAIC ESCAPE PARAMETER %%%
\subsubsection{Hydrodynamic escape parameter}

We examined the impact of two different values of the parameter $\lambda_p$ on the dynamics of escaping gas in the planetary atmosphere. Specifically, we analyzed models denoted by $L3$ (representative of a lower thermal energy compared to the fiducial model)and models denoted by $L{1.5}$ (representative of a higher thermal energy). 

When the thermal energy is decreased, the pressure $P_p$ of the launching wind is subsequently lowered as shown by Eq\,\ref{eq:press_p}. As a result, the speed of sound $c_s = \sqrt{\gamma P_p / \rho_p}$ is reduced, with $c_s \propto \lambda_p^{-1/2}$. The wind accelerates more slowly and consequently, the sonic surface is located at a greater distance from the planet, scaling as $r_s \propto \lambda_p$. Slower outflow leads to smaller cavities of planetary wind and narrower Doppler-shifted helium 1083~nm spectral lines.

Our investigation further revealed that the slope of the blueshift in Fig.\,\ref{fig:Vel_shift} is steeper for the $L1.5$ models compared to $L3$. At most degrees of anisotropy, the cooler outflows of case $L$3 (with $\lambda_p=3$) yield smaller magnitude of lower $v_{\rm shift}$ than model $A$. By contrast, the  hotter outflow of the $L$1.5 models leads to larger maximum blueshift before it saturates around model $L1.5_{40\%}$. We also observe that the stronger planetary outflows at smaller values of $\lambda_p$ require a higher degree of anisotropy to produce a noticeable fallback of material. 

\subsection{Constant planetary atmospheric density versus constant planetary mass-loss rate}\label{sec:const_m_d_p}

In our prior analyses, we maintained a constant initial density at the planet's surface to launch the wind, while modifying the degree of anisotropy. This approach resulted in a gradual decrease in the planetary mass-loss rate due to the reduced pressure on the night side, as outlined in Appendix \ref{sec:app_mass_loss} and Fig.\,\ref{fig:app_mass_loss}. We have opted for this approach to gain a deeper understanding of the simulations without introducing additional parameters that might add complexity to the analysis. Nevertheless, in order to untangle the impact of the varying planetary mass-loss rate $\dot{m}_p$, we also explored the behavior of the helium line properties for a constant $\dot{m}_p$ across all degrees of day-night anisotropy. This adjustment was achieved through a rescaling of the overall simulation density, as described in Sect\,\ref{sec:app_rescaling}, to match the planetary mass-loss rate of the isotropic case $A_{100\%}$.

\begin{figure}[h!]
    \centering
    \includegraphics[width=.499\textwidth]{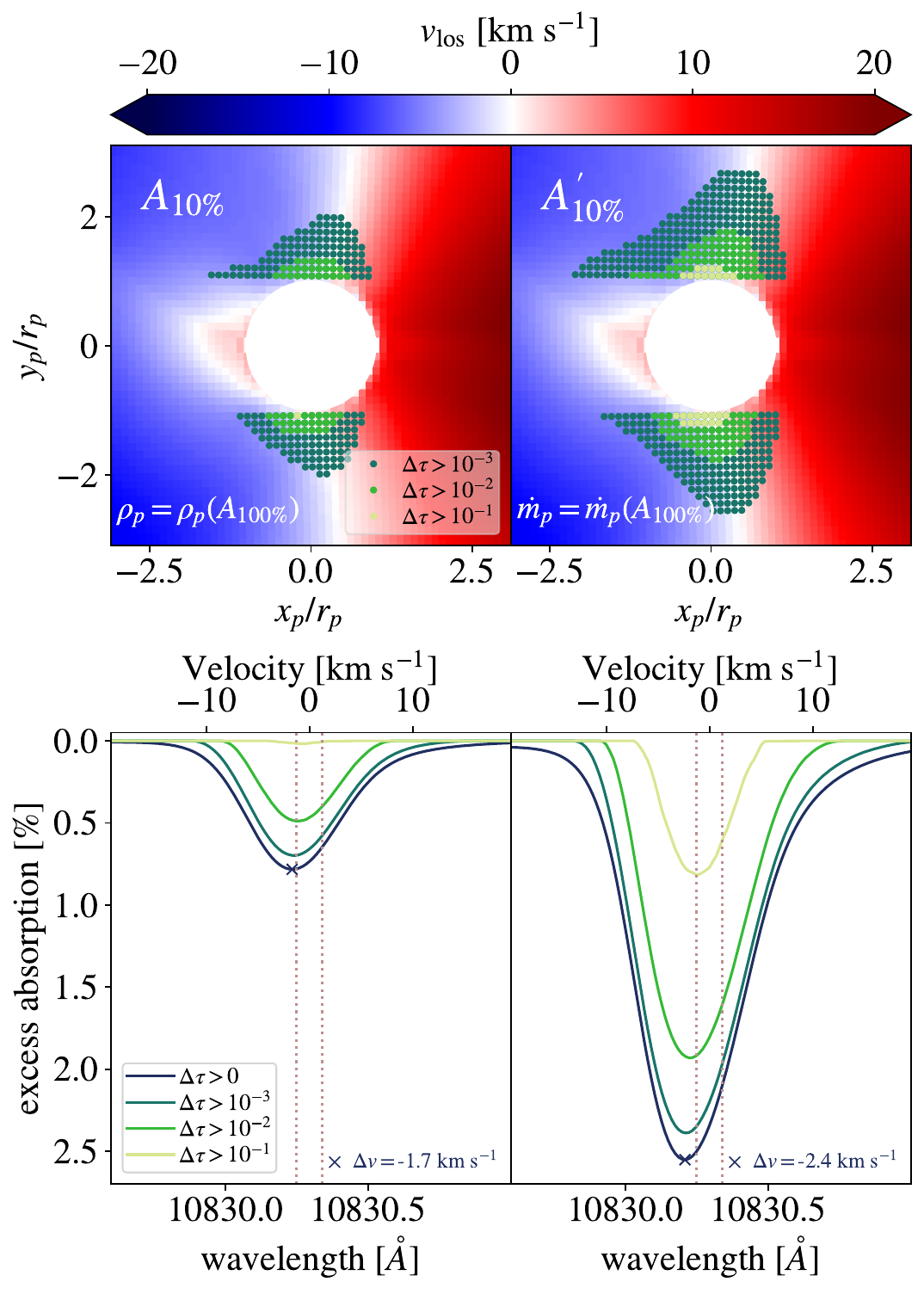}
    \caption{Comparative analysis of constant initial density vs. constant planetary mass-loss rate (same as Fig.\,\ref{fig:Blueshift}). The left panel depicts the scenario in which a uniform initial density ($\rho_p$) around the planet was maintained across varying degrees of day-night anisotropy, as previously discussed in our analysis. In contrast, the right panel shows the outcome of maintaining a consistent planetary mass-loss rate ($\dot{m}_p$) by aligning it with the values of the isotropic case $A_{100\%}$.}
    \label{fig:const_mdotp}
\end{figure}

In Fig.\,\ref{fig:const_mdotp}, we present a comparison of these two approaches: the left panel depicts the scenario with a constant initial density, as discussed in our previous analysis, while the right panel shows the situation with a constant planetary mass-loss rate. As a consequence of the density rescaling in the latter case, the density surrounding the planet increased, leading to a greater number of cells contributing to line formation. Consequently, the spectral line exhibited increased depth and equivalent width compared to the case of constant initial density. Furthermore, we observed that the cells responsible for line formation were located at higher altitudes in the right panel. As a result, the blueshift of the helium line is higher with $\Delta v = - 2.4$ km s$^{-1}$, in contrast to the $\Delta v = - 1.7$ km s$^{-1}$ observed in the former case for model $A_{10\%}$.

To compare the mid-transit velocity shift behavior with increasing degree of day-night anisotropy, we added model group $A'$ to Fig.\,\ref{fig:Vel_shift} (black dotted line). This model group shows a trend similar to that of the fiducial model group $A$. However, as day-night anisotropy increases, the blueshift of the helium line in $A'$ becomes slightly more prominent compared to $A$. This is attributed to the helium line's formation process occurring at higher altitudes, as discussed previously. A spectral time series depicting the excess absorption, akin to Fig.\,\ref{fig:Excess}, is presented in Fig.\,\ref{fig:Excess_A'}.

%-----------------------------------------------------------------
\section{Discussion} \label{sec:Discussion}

\subsection{Revealing day-night anisotropy}

\begin{figure*}
    \centering
    \includegraphics[width=\textwidth]{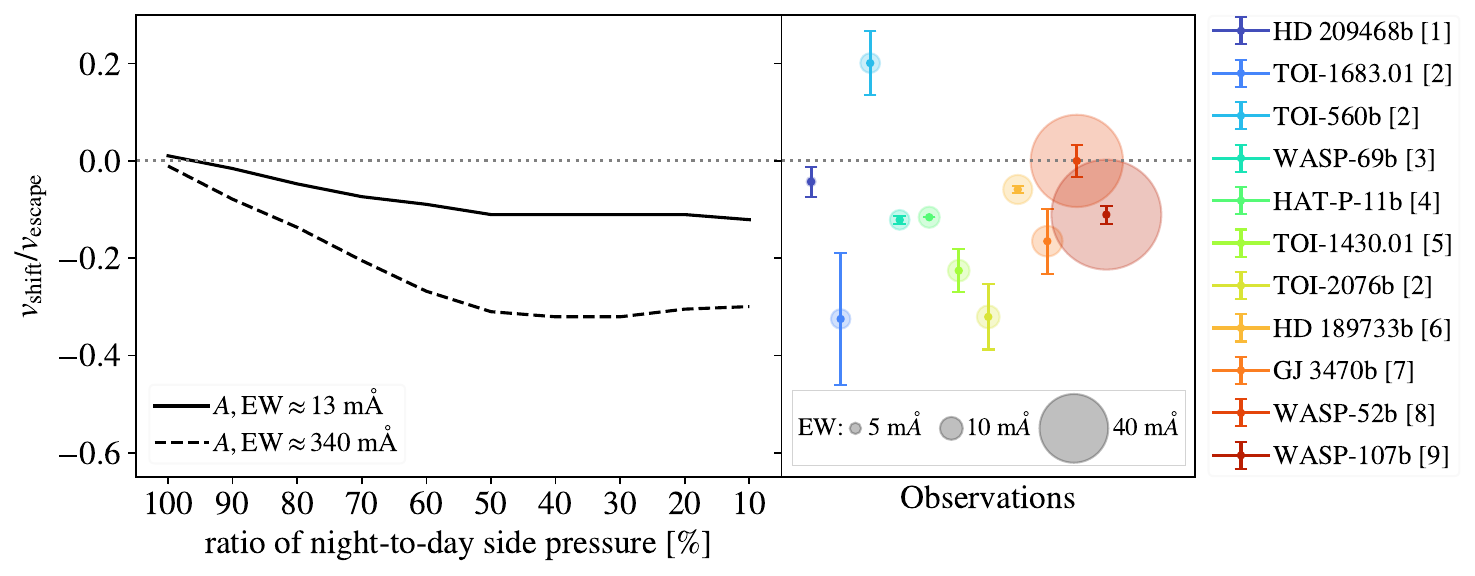}
    \caption{Mid-transit velocity shift of the strong (red) component of the helium 1083~nm line shown as a function of the night-to-day side pressure ratio, normalized by the escape velocity at the planet's surface. Here, we scale the density of the snapshots in a way that ensures the resulting spectra attain the same equivalent width (EW) \emph{(left)}. The observational measurements of the helium velocity shift during mid-transit in units of the individual escape velocity of the observed planet. The marker size corresponds to the strength of the measured EW \emph{(right)}. The dotted line shows the separation between positive and negative velocity shifts. By adjusting our simulations to match the properties of a specific planet, including the equivalent width (EW), we could effectively constrain the day-night anisotropy of the planet's atmosphere. No velocity shift uncertainties were provided for HAT-P-11b. [1] \cite{alonso-floriano_he_2019},
    [2] \cite{zhang_detection_2023},
    [3] \cite{nortmann_ground-based_2018},
    [4] \cite{allart_spectrally_2018},
    [5] \cite{2023A&A...677A..56O},
    [6] \cite{salz_detection_2018},
    [7] \cite{palle_he_2020},
    [8] \cite{kirk_kecknirspec_2022},    
    [9] \cite{kirk_confirmation_2020}.
    }
    \label{fig:Vel_shift_observations}
\end{figure*}

The transport of heat between the day side and the night side in tidally locked exoplanets, and the resulting day-night temperature gradients, have been extensively studied over the years. However, most of the work has been focused on investigating the energy circulation in the lower atmosphere, close to the planet's photosphere, while the uppermost atmospheric layers, such as those probed by the helium 1083~nm line, still remain largely underexplored.

3D atmospheric models of hot Jupiters revealed the formation of an equatorial jet at pressures around 1 bar. This jet leads to a displacement of the temperature maximum away from the substellar point in the direction of the planet’s rotation \citep[e.g.,][]{showman_atmospheric_2002, parmentier_3d_2013, showman_equatorial_2011}. The existence of this hotspot offset has been confirmed by phase curve observations in the infrared \citep[e.g.,][]{knutson_map_2007, stevenson_thermal_2014, bell_comprehensive_2021, may_new_2022}.

At pressure levels lower than about 1~mbar, 3D atmospheric models find that the wind structure changes to a flow directed from the day side to the night side \citep[][]{showman_atmospheric_2009, rauscher_three-dimensional_2010, heng_atmospheric_2011, miller-ricci_kempton_constraining_2012}. The pressure level at which this qualitative transition in the atmospheric flow occurs depends on the molecular weight of the atmosphere, with heavier molecules causing the transition to occur deeper within the atmosphere \citep{zhang_effects_2017}. The details of the day-to-night wind pattern depend on the treatment of the magnetic effects in the model atmosphere \citep[e.g.,][]{miller-ricci_kempton_constraining_2012, beltz_magnetic_2022}, but they are broadly consistent with spectroscopic observations, which find net blueshifts of a few km s$^{-1}$ \citep[e.g.,][]{Snellen_2010, nortmann_ground-based_2018, salz_detection_2018}.

Extending the predictions of 3D atmospheric models to even lower pressures is not straightforward. \cite{miller-ricci_kempton_constraining_2012} found that at pressures lower than 1~mbar, the winds can get stronger or weaker depending on the treatment of the magnetic drag. Recently, \cite{beltz_magnetic_2022} used the active magnetic drag treatment \citep{rauscher_three-dimensional_2013}, which calculates the local drag timescale throughout the atmosphere, and found that this approach most strongly affects the upper atmosphere, where the winds mainly flow from the day side to the night side over the poles. Observations of extended atmospheres, such as those made using the helium 1083\,nm line, could thus provide valuable constraints on these models.

%2. Return to the main question you posed early on in the abstract, and answer it.
Understanding the efficiency of heat distribution, both from the day to the night side and vertically from the photosphere to the upper low density regime probed by He~1083~nm observations, would be valuable in gaining deeper insights into the atmospheric dynamics. The question we aim to address in this study is whether we can infer the day-night temperature gradient from high-resolution helium 1083~nm observations, assuming the presence of day to night side winds at high altitudes. Our findings reveal a correlation between the velocity shift of the He 1083~nm triplet and the degree of day-night anisotropy, offering the potential to constrain the latter through careful analysis.

Therefore, for a particular planet, if we assume that the sole driver of the helium triplet velocity shift is day-night temperature gradient, we could simultaneously constrain three free parameters: the planetary mass loss rate, characteristic temperature of the outflow, and degree of day-night anisotropy. To construct these constraints requires a high-resolution transit spectrum with a resolved He 1083~nm line profile.  To apply this approach practically,  the spectral energy distribution (SED) of the star is needed to calculate the population of metastable helium in the gas.  Though these three parameters are linked, we can make some generalizations about their impact on the transmission spectrum. The temperature of the outflow informs the line's Doppler-broadened width \citep{linssen_constraining_2022}. At a given temperature, the line equivalent width is determined by the density of the outflow, or the planetary mass-loss rate. A 1D Parker wind approach can also provide a good starting point for 3D modeling \citep[e.g.,][]{oklopcic_new_2018, linssen_constraining_2022}. Finally, as we have shown in this work, the velocity shift (blue or red) is informed by any day-night anisotropy of the planetary upper atmosphere. By combining these steps, we can constrain the day-night anisotropy as a free parameter by assigning the velocity shift to the model correlation. 

We anticipate that one aspect of the constraints one could derive would be a limiting value of the anisotropy parameter. Our study has provided insights into the impact of a fallback of gas on the night side, as described in Sect\,\ref{sec:Fallback}. In such a scenario, the blueshift saturates at higher degrees of day-night anisotropy. It becomes challenging to differentiate between a very strong degree of anisotropy and an intermediate degree, as the blueshift reaches a maximum value. This saturation effect poses a limitation on what can be  distinguished observationally based  on the magnitude of the blueshift.

%3. Explain how your answer is supported by your data, model, figures, derivations, results.
As a less planet-specific overview than the approach outlined above, in Fig.\,\ref{fig:Vel_shift_observations}, we present a comparison between the helium line shifts observed in our simulations from model group $A$ and those reported in helium transit observations. We note that the simulations are not scaled to the properties of the particular systems in question, which would require simulating each system separately. However, the relative robustness of our results to model variations (e.g., as seen in Fig.\,\ref{fig:Vel_shift}) motivates this comparison. To ensure comparability, we express the observed velocity shift for each planet in terms of the planet's escape velocity at the surface. The right side of the figure displays the observational results, ordered from left to right according to increasing equivalent width. Many of them exhibit a blueshift that covers a velocity range that aligns with the outcomes of our study. On the left side, we present model group $A$ scaled to two different EW values. The solid line corresponds to the smaller value of $13~\rm m\AA$, while the dashed line represents the larger value of $340~\rm m\AA$. To achieve a higher EW, we scaled the snapshots to a higher density, effectively placing the line-forming region at higher altitudes with increased wind speeds, leading to a higher velocity shift. %It is important to note that the scaling is arbitrary and can be adjusted to match any desired equivalent width. This demonstrates the versatility of our approach in accurately modeling the helium line profiles for different planetary conditions.

In Fig.\,\ref{fig:Vel_shift_observations} we note that many of the observed systems show a velocity blueshift on the order of 10-40\% of the planetary escape velocity. These sorts of velocity shifts appear compatible with the magnitude of day-to-night flows seen in our simulations. However, our simulations predict a correlation between the magnitude of the velocity shift and the line equivalent width (all else being equal), which is not seen when looking across the observed systems. Future work with the planet-specific simulation approach described above will probe how well the day-night anisotropy model applies to particular systems. 

\subsection{Limitations}

% CASE C %
When interpreting the helium line shift observed in transmission spectroscopy of evaporating gas giants, we encounter two types of constraints. Those that affect the blueshift when day to night side winds are present, which have been discussed above in the context of jointly fitting outflow parameters, and those that cause a blueshift by an entirely different physical process, unrelated to the day-night anisotropy. 

The latter includes a strong stellar wind relative to the planetary wind, as shown by our model group C. Even using an isotropic boundary condition for the planetary wind results in a blueshift of the helium line, as shown in Fig.\,\ref{fig:Blueshift}. However, the work by \cite{macleod_stellar_2022} has demonstrated that, in such scenarios, a distinct post-transit signal emerges due to the planet's flow being considerably confined within a cometary trailing tail. The resulting asymmetric light curve can help to distinguish cases influenced by a strong stellar wind from those primarily driven by day-night anisotropy. %Furthermore, the large error bars observed in \ref{fig:Vel_shift} suggest that significant temporal variability of the spectroscopic feature may also serve as an indication of the presence of a strong stellar wind. This variability could potentially lead to uncertainties in the measurement of the velocity shift, and careful consideration should be given to account for the effects of temporal fluctuations when interpreting the observed data. 

% MAGNETICALLY_CONTROLLED OUTFLOW %
Furthermore, we did not study the effects of a planetary magnetic field and the extent to which they could impede the flow from the day to the night side. Previous magnetohydrodynamic simulations have suggested that the outflow from a dipole magnetic field would be confined to the polar regions, resulting in a double-tailed structure, with the opening angle decreasing as the field strength increases \citep{adams_magnetically_2011, trammell_hot_2011, trammell_magnetohydrodynamic_2014, carolan_effects_2021}. In a recent study by \cite{schreyer_using_2023}, 2D simulations of photoevaporation were performed for a configuration involving a strong magnetic field. The results indicate that the day-to-night side winds can be suppressed due to the presence of a planetary magnetic field, resulting in either a lack of net blueshift or even a detectable redshift in the helium line.

% RADIATION PRESSURE %
We note that the influence of 1083~nm radiation pressure on gas dynamics is ignored in this study. There are different models and approaches in the literature regarding the treatment of metastable helium in gas giant atmospheres, with some models considering it as collisionless particles \citep[e.g][]{allart_high-resolution_2019}, while others treat it as a separate fluid species in a multi-fluid approach \citep[e.g][]{khodachenko_impact_2021}. The latter model finds substantial differential velocity for metastable helium as it is accelerated by radiation between collisions, which differs from the results presented in the study by \cite{wang_metastable_2021-1}. The estimation of the short mean free path near the planet, discussed in \cite{macleod_stellar_2022}, suggests that the planetary outflow probed by the helium line is in a collisional regime for observed mass-loss rates, making the use of hydrodynamic simulations appropriate.  However, it is important to note that radiation pressure could still act as a radial force applied from the star -- much like the action of a stellar wind, collimating and confining planetary outflow in particularly strong circumstances. 

% STELLAR REDUCTION %
Finally, we emphasize the utility of having a long baseline for helium observations in order to obtain precise measurements of velocity shift. As revealed by recent studies of \cite{zhang_detection_2023} and \cite{gully-santiago_large_2023}, the escaping atmosphere can extend a significant fraction of the orbital phase -- hundreds of planetary radii. If out-of-transit spectra used to establish a baseline have the signature of planetary outflow kinematics, this can introduce biases in determining the velocity shift measured at mid-transit, as exemplified by \cite{spake_posttransit_2021}.

%-----------------------------------------------------------------
\section{Summary and conclusions} \label{sec:Summary}

% OBJECTIVE
The objective of this study was to investigate the gas dynamic effects of day-to-night temperature contrasts on the escaping atmosphere of a tidally locked planet, with an emphasis on analyzing the effects on the helium 1083~nm triplet.

% METHODS
By combining hydrodynamic simulations and radiative transfer post-processing, we modeled the spectra of the metastable helium triplet for planets in transiting configurations. Our simulations incorporate parameterized treatments of both planetary and stellar winds, allowing us to investigate the influence of wind properties on the observable characteristics of evaporating atmospheres in a wide range of parameters. We conducted a parametric study, varying the stellar mass-loss rate, orbital distance, and the ratio of thermal and gravitational potential energies (i.e., the hydrodynamic escape parameter, $\lambda_p$) associated with the planetary wind. The key findings of our study are:  

% MAIN FINDINGS
  
\begin{enumerate}
    \item With increasing the day-night anisotropy, there is a general trend toward narrowing of the helium line and an increase in the net blueshift of the line centroid.
    \item  The velocity shift of the line is contingent on the line forming altitude. A higher planetary mass-loss rate causes the line to form at higher altitudes, where the wind attains higher speeds. Consequently, this leads to a more pronounced velocity shift in the observed spectra.    
    \item We have identified a critical point of anisotropy, at which the blueshift stagnates, indicating that a saturation of the day-night anisotropy is reached. This saturation is attributed to turbulent flows generated by the outflow material falling back onto the night side of the planet.
    \item  The velocity shift of the helium line is affected the stellar wind strength and the presence of turbulent flows, potentially leading to time variations in the shift.
\end{enumerate}

% IMPLICATIONS
Assuming the day-night temperature gradient as the primary cause of the observed blueshifts in the He-1083~nm line, we can make use of the correlation found in Fig.\,\ref{fig:Vel_shift} to constrain the temperature gradient. This requires a simulation catered to the particular properties of an exoplanetary system. Because of the sensitivity of the line shift to the formation altitude, it is important to scale the density of the simulation appropriately to match the observed equivalent width.

% LIMITATIONS
However, we emphasize that day-night temperature gradients are not the only possible cause of blueshifts in He 1083~nm transit spectra. One other physical process of potentially high importance is the impact of a strong stellar wind, which could induce a blueshift in the helium line, regardless of the presence of day-night anisotropy. Previous research by \cite{macleod_stellar_2022} has shown that examining the light curve for asymmetry can be helpful in identifying this scenario. Additionally, processes not accounted for in our model, such as magnetic drag, magnetically controlled outflows, and radiation pressure, could also impact the helium velocity shift. 

The saturation of the blueshift in scenarios with falling back material on the night side suggests that it may be challenging to distinguish between intermediate and high day-night anisotropy cases. We find that consideration should be given to the effects of time variability when interpreting the observed line shift. This does not only concern the variability of the stellar UV flux, which affect the population of metastable helium in the atmosphere. Our models indicate that high variability implies that the line forming region includes turbulent flow, either because of eddies caused by fallback or wind-wind collisions.

%------------------------------------------------------------------

\begin{acknowledgements}
      We appreciate the helpful discussions with D. Linssen and C. Dominic. We thank the referee for their insightful comments, and we thank SURFsara (\url{www.surfsara.nl}) for their support in using the Lisa Compute Cluster.
\end{acknowledgements}

%-------------------------------------------------------------------

\bibliography{lib}{}
\bibliographystyle{aa}
%-------------------------------------------------------------------

\begin{appendix}

\section{Density rescaling} \label{sec:app_rescaling}

In hydrodynamic simulations conducted using codes such as \texttt{Athena++}, the physical quantities in a single snapshot can be manipulated without disrupting the flow patterns. This capability stems from the inherent scale invariance of the hydrodynamic equations utilized in these simulations, as they are solved numerically using discrete time steps. By uniformly increasing the pressure and density across the entire simulation domain, the relative values and ratios between pressure and density remain unaltered. Consequently, the flow patterns, which are determined by the gradients and variations in pressure and density, remain unchanged.

\begin{figure}
    \centering
    \includegraphics[width=0.49\textwidth]{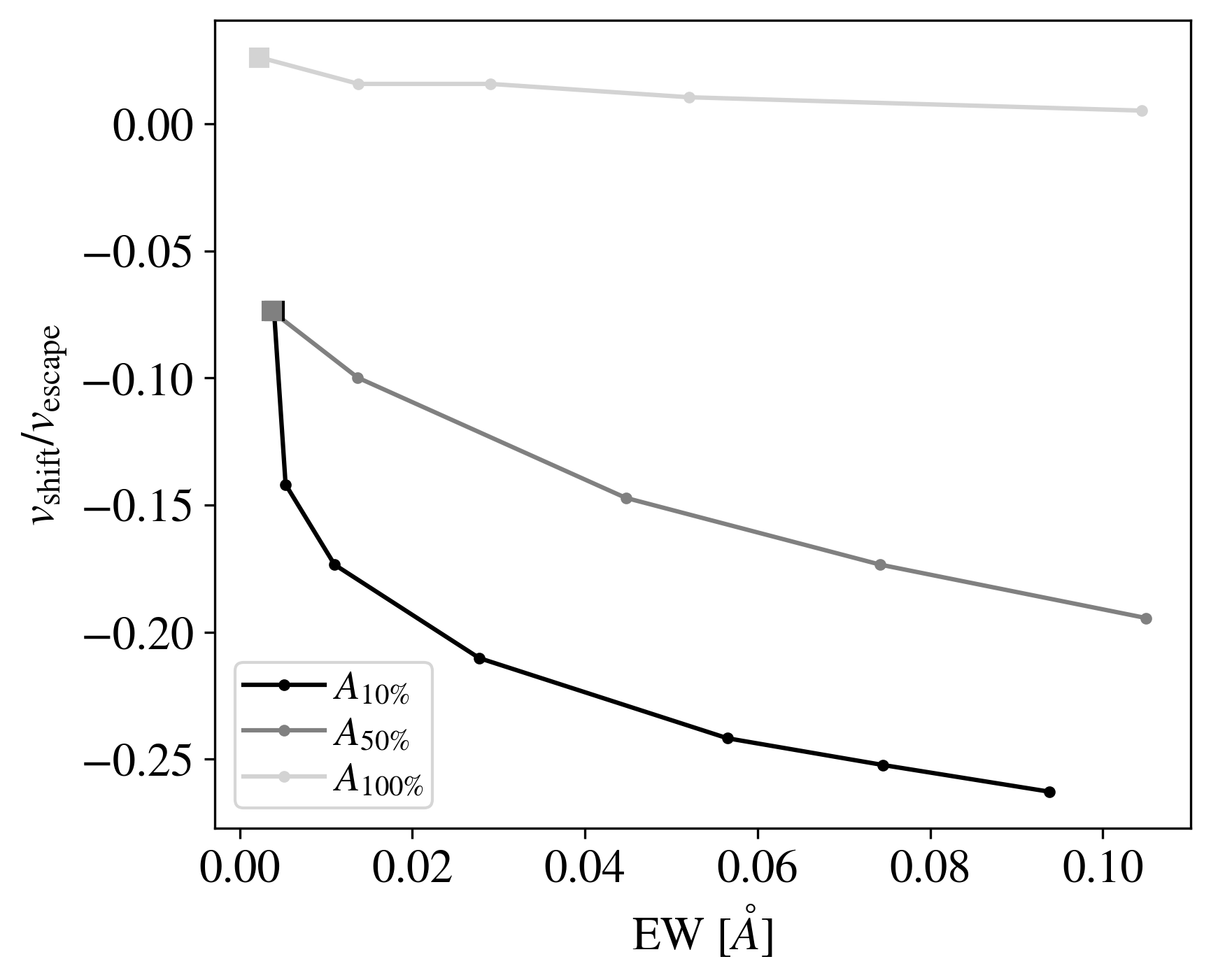}
    \caption{Mid-transit velocity shift of the strong (red) component of the helium 1083~nm line as a function of the equivalent width (EW), normalized by the escape velocity at the planet’s surface. The points marked by squares indicate the scaling value 1.0, which was used for the previous analysis in this paper. When the density is increased, the mass-loss rate of the planet also increases, leading to a larger equivalent width and a higher blue shift in the helium line.
    }
    \label{fig:app_EW+Shift}
\end{figure}

Taking advantage of this rescaling capability, we investigated the behavior of the blueshift exhibited by the helium line within a higher density environment surrounding the planet. Increasing the density is equivalent to increasing the mass-loss rate of the planet. In Fig.\,\ref{fig:app_EW+Shift}, we illustrate the line properties of the metastable helium triplet; specifically, the mid-transit velocity shift as a function of the equivalent width, for different density scalings. A clear trend is observed, where the velocity shift increases with an increasing equivalent width. The curves corresponding to individual models, ranging from isotropic wind to anisotropic wind, do not intersect. This implies that there is no degeneracy among different degrees of anisotropy. Consequently, the results presented in Fig.\,\ref{fig:Blueshift} would undergo an offset shift along the velocity axis in the case of different degrees of anisotropy.

An increase in gas density also leads to an increase in the number density of metastable helium. Consequently, the extent of the region where line formation occurs expands. In the context of day-night anisotropy, it is notable that the line predominantly forms on the night side, where the gas moves in the direction of the observer. At higher altitudes, the gas is more accelerated and as a result, the line exhibits a more pronounced blueshift. Moreover, the equivalent width of the line increases due to a greater contribution from the additional gas involved in line formation, leading to an increase in absorption depth.

\section{Planetary mass-loss rate} \label{sec:app_mass_loss}

The net mass-loss rate of the planet is calculated by determining the rate of change of its mass through the gradient of a passive scalar with respect to time. This gradient quantifies the variations in mass over small time intervals. Dividing this gradient by the corresponding time intervals yields the rate of change of mass per unit time. To account for the mass loss attributed to the planet's wind, which is already measured in the simulation, we subtract this contribution from the rate of change of mass. Consequently, the resulting value represents the net mass-loss rate of the planet, encompassing all mass loss mechanisms. Finally, we take the time average $\langle \dot{M}_p \rangle$ for $t>10^{6}$~s, when the quasi-steady state is reached. 

\begin{figure}
    \centering
    \includegraphics[width=0.43\textwidth]{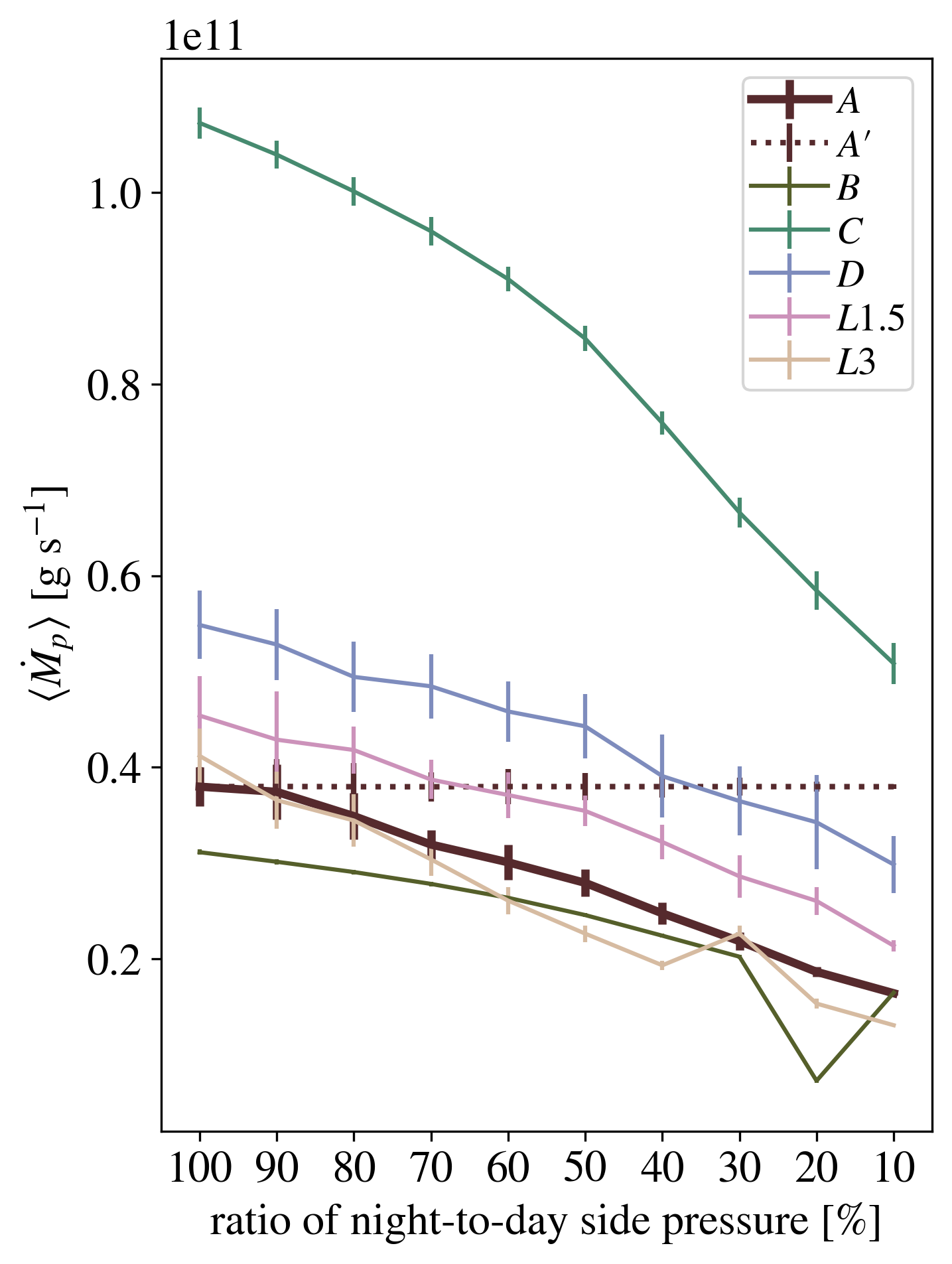}
    \caption{Time-averaged planetary mass-loss rates. The error bars indicate the standard deviation. The introduction of an anisotropic surface lowers the pressure, leading to a decrease in the mass-loss rate for increasing day-night anisotropy. Deviations for ratios $<50\%$ can be attributed to falling back material on the night side.}
    \label{fig:app_mass_loss}
\end{figure}

In Fig.\,\ref{fig:app_mass_loss}, we present the time-averaged planetary mass-loss rates with increasing day-night anisotropy. The mass-loss rates across all models exhibit a twofold decrease, with the exception of model group $A'$ intentionally scaled to maintain a constant $\dot{m}_p$. Firstly, the introduction of an anisotropic surface lowers the pressure, thereby reducing the driving force behind the wind, leading to a decrease in the mass-loss rate. Since the pressure is uniformly decreased in each model, the slope of the mass-loss rates as a function of the night-to-day side pressures should be the same across all models. Secondly, the presence of gas falling back onto the night side is expected to contribute to a decrease in the planetary mass-loss rate. Figure \ref{fig:app_mass_loss} demonstrates that, in general, the mass-loss rates exhibit a similar negative slope going from isotropic to anisotropic conditions. However, model $B$ and $L3$ show a significant deviation at strong anisotropic conditions, while models $A$ and $C$ show a negative increase in slope. We observe that the extent of the deviations correlate with the occurrence and strength and of the falling back material on the night side. The remaining models, namely $D$, and $L1.5$, do not exhibit significant variations.

\section{Numerical effects on radiative transfer results} \label{sec:app_numerical}

\begin{figure}
    \centering
    \includegraphics[width=.47
\textwidth]{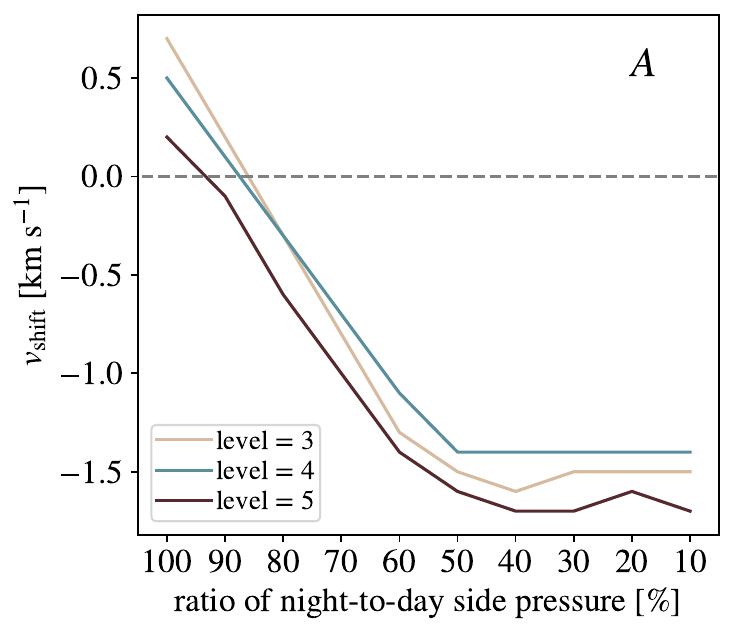}
    \caption{Mid-transit velocity shift of the dominant (red) component in the helium 1083~nm triplet as a function of the night-to-day side pressure;  post-processed with different spatial resolution levels of the simulation for model group $A$. the overall numerical uncertainty of the velocity shift is in the range of $\sim 0.5$ km~s$^{-1}$.}
    \label{fig:app_resolution}
\end{figure}

Figure \ref{fig:app_resolution} illustrates the impact of the spatial resolution during the radiative transfer post-processing step on the mid-transit velocity shift of the helium line. In the case of the isotropic model $A_{100\%}$, increasing spatial resolution leads to a convergence of the velocity shift toward zero (0.7, 0.5, 0.2 km~s$^{-1}$ for levels 3, 4, and 5, respectively). However, this monotonic trend is not evident in all the anisotropic models due to their more complex flow patterns. However, the curves corresponding to different resolution levels follow the same trend and span a range of approximately $\sim 0.5$ km~s$^{-1}$, which we estimate reflects the overall numerical uncertainty associated with the velocity shifts measured in this study.

Furthermore, we investigated the effects of the convergence of the mid-transit velocity shift with different total numbers of rays ($N_{\rm radial}$ = 30, 50, 100) and the step size coefficient along the rays ($f_l$ = 0.01, 0.5, 0.1). The differences in the blueshift for the tested parameters were not significant. For the results shown in this work, we used for the radiative transfer post-processing step the spatial resolution level 4, alongside $N_{\rm radial} = 100$ and $f_l$ = 0.1.

\section{Additional figures}
\begin{figure*}
    \centering
    \includegraphics[width=.95\textwidth]{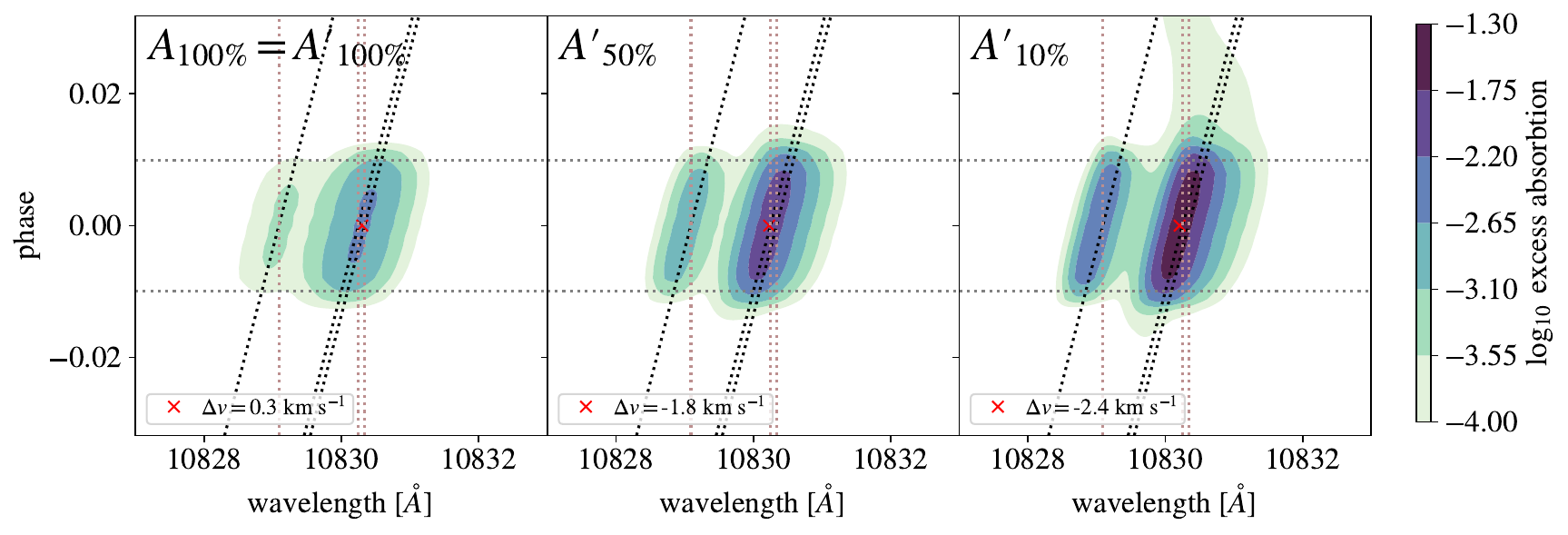}
    \caption{
    Analogous representation of metastable helium absorption in the stellar frame as shown in Fig.\,\ref{fig:Excess}, but for model group $A^{\prime}$, where the planetary mass-loss rate matches that of model $A_{100\%}$. Similarly to model group $A$, where the density is the same across all models, the line becomes  narrower and blueshifted as the day-night anisotropy increases. As discussed in Sect\,\ref{sec:const_m_d_p}, in the case of constant planetary mass-loss rate, the line forms at higher altitudes which leads to a higher blueshift and absorption.} 
    \label{fig:Excess_A'}
\end{figure*}

Figure \ref{fig:Excess_A'} shows Fig.\,\ref{fig:Excess} for model group $A^{\prime}$. To complement this visual representation, Table \ref{tab:app_models} provides further details on line properties across all models. Additionally, it lists simulation measurements of both stellar and planetary mass-loss rates.

\begin{figure}
    \centering
 \includegraphics[height=.93\textheight]{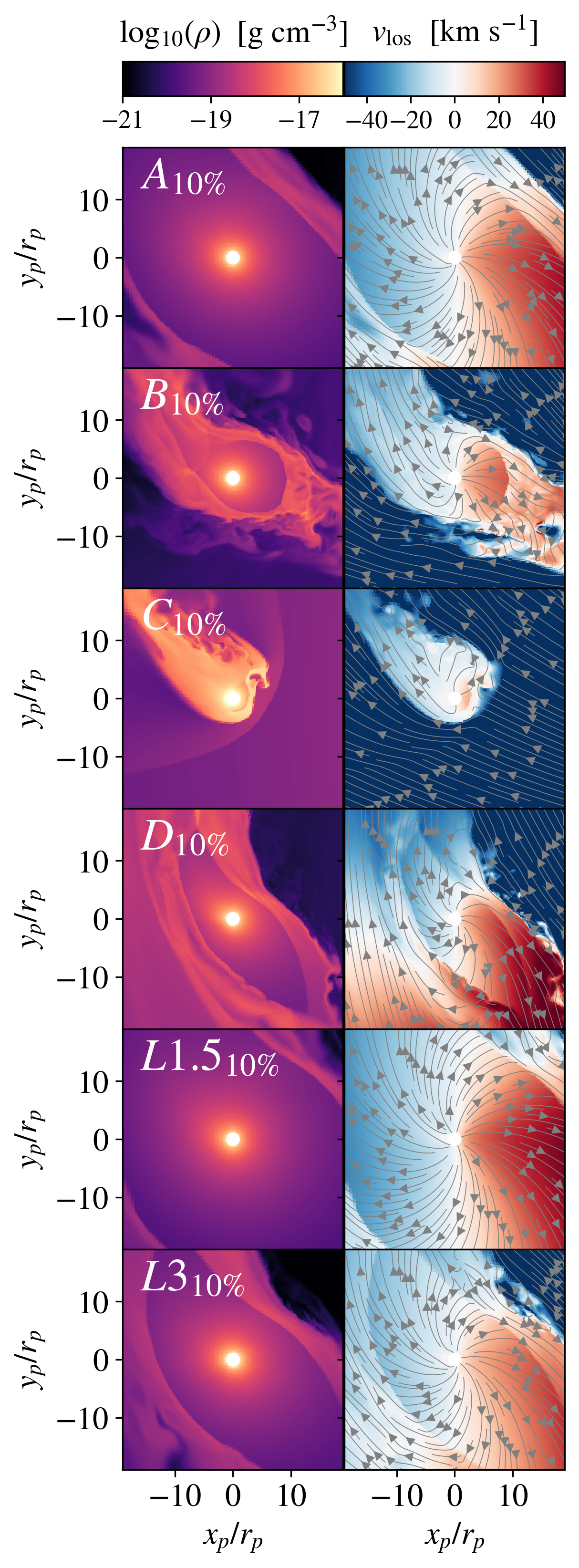}
    \caption{Anisotropic model overview of the density (left) and line-of-sight velocity (right) showing the varying morphology of the planet's vicinity. The flow directions are indicated by gray streamlines. The observer is located in the $-x$ and the star in the $+x$-direction.}
    \label{fig:app_overview}
\end{figure}

In Fig.\,\ref{fig:app_overview}, the density and line-of-sight velocity are displayed in the vicinity of the planet for all anisotropic models with \fpresT = 0.1. This figure illustrates how the shape and size of the cavity surrounding the planet change across the models studied in this work.

\begin{figure}
    \centering
    \includegraphics[width=.45\textwidth]{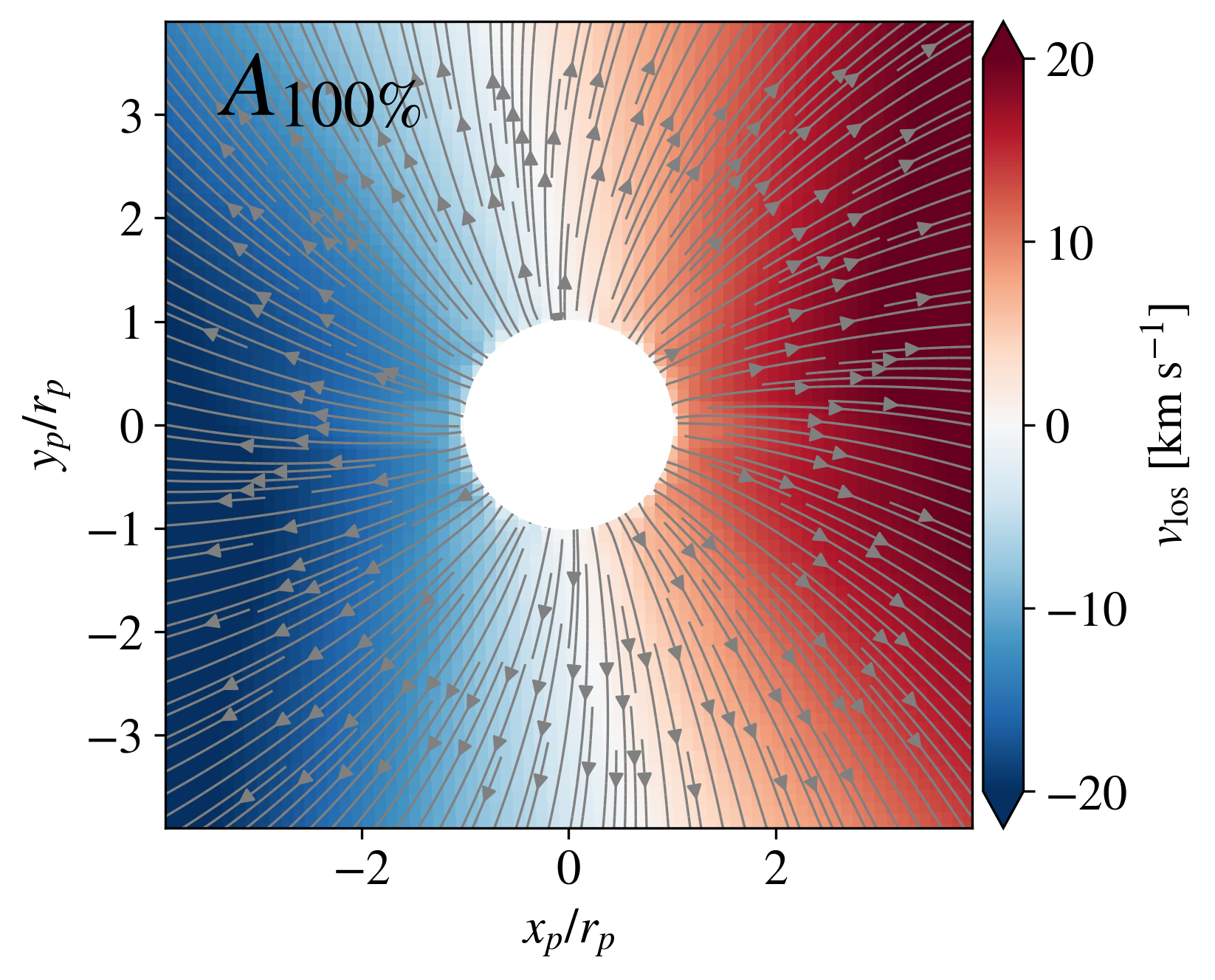}
    \caption{(Movie online) A sequence of evolved snapshots displaying the line-of-sight velocity around the planet, accompanied by over-plotted streamlines depicting the infalling material on the night side at high degrees of day-night anisotropy; with model $A_{100\%}$ representing the isotropic case and model $A_{10\%}$ exemplifying the most anisotropic scenario. Notably, the occurrence of fallback can be directly linked to the saturation of the blueshift shown in Figure \ref{fig:Vel_shift}.}
    \label{fig:app_movA}
\end{figure}

\begin{figure}
    \centering
    \includegraphics[width=.45\textwidth]{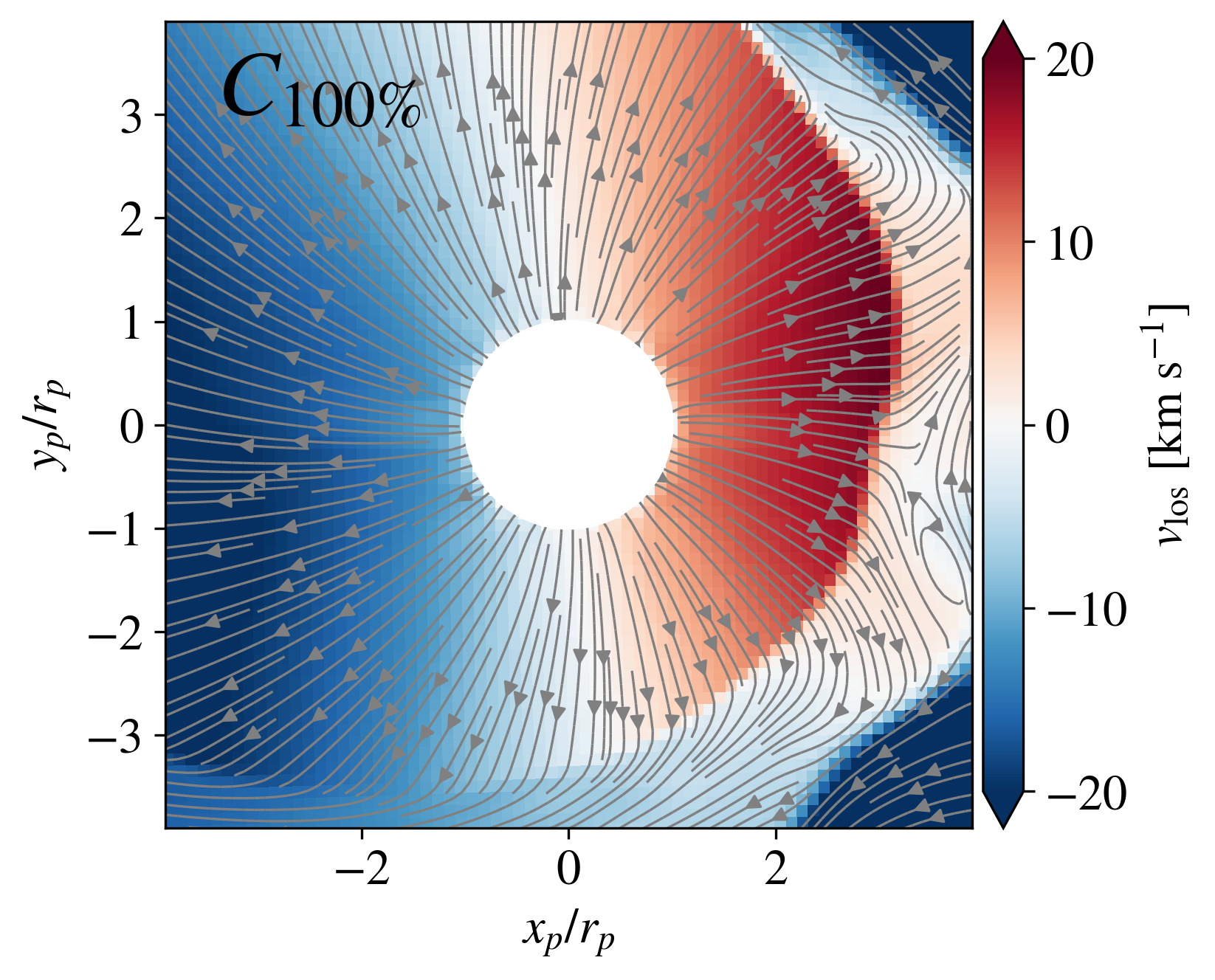}
    \caption{(Movie online) Same as Fig.\,\ref{fig:app_movA}, but showing model group $C$. It shows that the cavity around the planet dissipates on the night side of the planet as the planetary wind becomes too weak to form a shock front with increasing day-night anisotropy. Additionally, a striking observation is that the shocked regions now align with the helium line formation region. This alignment results in a significant temporal variability, as depicted in Figure \ref{fig:Vel_shift}.}
    \label{fig:app_movC}
\end{figure}

The online movies linked to Fig.\,\ref{fig:app_movA} and Fig.\,\ref{fig:app_movC} show snapshots of the line-of-sight velocity and streamlines in the vicinity of the planet for increasing day-night anisotropy. 

\begin{table*}[tbp]
\caption{Simulation results and helium line properties.}
\centering
    \begin{tabular}{lccccccccc}
    \toprule
         model & $\lambda_p$ & $a$ &  $\langle \dot{M_*} \rangle$ & $\langle \dot{M_p} \rangle$ & $\delta$ & $\delta_{\rm model} / \delta_{A_{100\%}}$  & EW & FWHM & $\Delta v $  \\
        \midrule
         & \ & [au] & [g~s$^{-1}$]  & [g~s$^{-1}$] & [$\%$] &  & [m$\AA$] & [$\AA$] & [km~s$^{-1}$] \\
        \midrule
        $A_{100\%} = A^\prime_{100\%}$ & 2 & 0.05 &  $7.70 \times 10^{10}$ &  $3.80 \times 10^{10}$  & 0.25  &  0.97  &  2.15  &  0.70  &  0.25 \\
        $A_{50\%}$ & 2 & 0.05 &  $8.19 \times 10^{10}$ &  $2.79 \times 10^{10}$  
        & 0.58  &  2.72  &  3.59  &  0.51  &  -1.64 \\
        $A_{10\%}$ & 2 & 0.05 &  $8.64 \times 10^{10}$ &  $1.63 \times 10^{10}$  
        & 0.78  &  4.90  &  3.86  &  0.40  &  -1.69 \\
        \midrule
        $A^\prime_{50\%}$ & 2 & 0.05 &  $ 1.11 \times 10^{11}$ &  $3.80 \times 10^{10}$ 
        &  0.95  &  4.17  &  6.06  &  0.51  &  -1.79 \\
        $A^\prime_{10\%}$ & 2 & 0.05 &  $ 2.00 \times 10^{11}$ & $3.80 \times 10^{10}$ 
        &  2.55  &  12.99  &  14.54  &  0.46  &  -2.39 \\
        \midrule
        $B_{100\%}$ & 2 & 0.05 & $8.79 \times 10^{11}$ &  $3.11 \times 10^{10}$   
        &  0.33  &  0.97  &  2.81  &  0.67  &  0.25 \\
        $B_{50\%}$ & 2 & 0.05 & $8.77 \times 10^{11}$ &  $2.46 \times 10^{10}$   
        & 0.61  &  2.72  &  4.03  &  0.52  &  -1.64 \\
        $B_{10\%}$ & 2 & 0.05 & $8.75 \times 10^{11}$ &  $1.65 \times 10^{10}$   
        &  0.86  &  4.53  &  4.81  &  0.43  &  -1.94 \\
        \midrule
        $C_{100\%}$ & 2 & 0.05 & $8.75 \times 10^{12}$ &  $1.07 \times 10^{11}$ 
        & 6.87  &  53.47  &  51.59  &  0.61  &  -2.70 \\
        $C_{50\%}$ & 2 & 0.05 & $8.71 \times 10^{12}$ &  $8.47 \times 10^{10}$  
        & 10.98  &  87.16  &  82.01  &  0.61  &  -4.78 \\
        $C_{10\%}$ & 2 & 0.05 & $8.71 \times 10^{12}$ &  $5.11 \times 10^{10}$   & 15.53  &  139.26  &  82.28  &  0.43  &  -2.70 \\
        \midrule
        $D_{100\%}$ & 2 & 0.03 & $5.73 \times 10^{10}$ & $5.44 \times 10^{10}$  
        &  0.11  &  0.36  &  1.04  &  0.75  &  -0.09 \\
        $D_{50\%}$ & 2 & 0.03 & $5.95 \times 10^{10}$ &  $4.43 \times 10^{10}$  
        &  0.25  &  1.08  &  1.71  &  0.53  &  -1.49 \\
        $D_{10\%}$ & 2 & 0.03 & $6.31 \times 10^{10}$ &  $3.00 \times 10^{10}$   
        &  0.33  &  1.96  &  1.82  &  0.42  &  -1.39 \\
        \midrule
        $L1.5_{100\%}$ & 1.5 & 0.05 &  $7.40 \times 10^{10}$ & $4.40\times 10^{10}$  
        &  0.12  &  0.37  &  1.30  &  0.84  &  0.20 \\
        $L1.5_{50\%}$ & 1.5 & 0.05 &  $7.87 \times 10^{10}$ & $3.54 \times 10^{10}$  
        &  0.25  &  0.54  &  2.03  &  0.65  &  -2.64 \\
        $L1.5_{10\%}$ & 1.5 & 0.05 &  $8.35 \times 10^{10}$ &  $2.12 \times 10^{10}$  
        &  0.45  &  1.95  &  2.62  &  0.47  &  -2.64 \\
        \midrule
        $L3_{100\%}$ & 3 & 0.05 &  $7.72 \times 10^{10}$ &  $4.15\times 10^{10}$ 
        &  1.33  &  7.85  &  8.25  &  0.51  &  0.25 \\
        $L3_{50\%}$ & 3 & 0.05 &  $8.55 \times 10^{10}$ &  $2.24\times 10^{10}$ 
        &  2.08  &  15.14  &  10.02  &  0.40  &  -0.74 \\
        $L3_{10\%}$ & 3 & 0.05 &  $8.87 \times 10^{10}$ &  $1.31\times 10^{10}$   
        &  1.88  &  14.40  &  8.09  &  0.35  &  -1.69 \\
    \bottomrule
    \end{tabular}
    \\
    \raggedright
        Notes. Time-averaged stellar $\langle \dot{M_*} \rangle$  and planetary $\langle \dot{M_p} \rangle$  mass-loss rates measured from simulation results, as well as line properties of the synthetic metastable helium spectrum at phase 0, where $\delta$ = absorption depth, EW = equivalent width, FWHM = full width at half maximum, $\Delta v$ = velocity shift of the line centroid. These values are extracted from a single spectrum of a representative simulation snapshot.
    \label{tab:app_models}
\end{table*}

\section{Data and software availability}
Simultaneously with the publication of this article, we have released the necessary software and data to replicate the results and analysis. The hydrodynamic model snapshots (in \texttt{Athena++} hdf5 format), as well as 3D post-processed species number densities (hdf5 format), and the synthetic spectra (in ascii format) of models $A_{100\%}$, $A_{50\%}$, and $A_{10\%}$ can be accessed on Zenodo at \href{https://zenodo.org/doi/10.5281/zenodo.10025849}{10.5281/zenodo.10025849}. Furthermore, we included the radiative transfer post-processing software used in our study, the modified \texttt{Athena++} setup files, and the code necessary to replicate the presented figures.

\end{appendix}

\end{document}